\documentclass[twocolumn,onecolappendix]{aastex63}
\received{Feb 1, 2022}
\revised{May 10, 2022}
\accepted{\today}
\submitjournal{ApJ}

\def\be{\begin{equation}}
\def\ee{\end{equation}}
\def\CF3{{\sc cosmicflows-3}}
\usepackage{rotating}
\usepackage{multirow}
\usepackage{CJKutf8}
\usepackage{graphicx}	
\usepackage{amsmath}	
\usepackage{amssymb}	
\usepackage{hyperref}
\usepackage{newtxtext,newtxmath}

\usepackage[T1]{fontenc}
\usepackage{ae,aecompl}

\def\papI{\href{https://academic.oup.com/mnras/article/487/4/5209/5498307}{Paper I}}

\def\papII{\href{https://academic.oup.com/mnras/article/487/4/5235/5513479}{Paper II}}
 
\usepackage{color}

\shorttitle{The Redshift-Space Momentum Power Spectrum III}
\shortauthors{Qin et al.}
 
\graphicspath{{./}{figures/}}

\begin{document}

\title{The Redshift-Space Momentum Power Spectrum III: measuring the growth rate from the SDSSv survey using auto- and cross- power spectrum of the galaxy density and momentum fields}

\correspondingauthor{Fei Qin}
\email{qin@cppm.in2p3.fr}

\author[0000-0001-7950-7864]{Fei Qin}
\altaffiliation{Aix-Marseille University, CNRS/IN2P3,  CPPM, Marseille, France.}
\affiliation{Aix-Marseille University, CNRS/IN2P3, CPPM, Marseille 13288, France}
\affiliation{School of Physics, Korea Institute for Advanced Study, Dongdaemun-gu,  Hoegiro 85, Seoul 02455, Republic of Korea}

\author[0000-0002-1081-9410]{Cullan Howlett}
\affiliation{School of Mathematics and Physics, The University of Queensland, Brisbane, QLD 4072, Australia}

\author[0000-0002-7464-2351]{David Parkinson}
\affiliation{Korea Astronomy and Space Science Institute, Yuseong-gu, Daedeok-daero 776, Daejeon 34055, Republic of Korea}

\begin{abstract}
The large-scale structure of the Universe and its evolution over time contains an abundance of cosmological information. 
One way to unlock this is by measuring the density and momentum power spectrum from the positions and peculiar velocities of galaxies, and fitting  the cosmological parameters from these power spectrum.
In this paper, we will explore the cross power spectrum between  the  density and momentum fields of galaxies. We derive the estimator of the density-momentum cross power spectrum multipoles. The growth rate of the large-scale-structure, $f\sigma_8$ is measured from fitting the combined density monopole, momentum monopole and cross dipole power spectrum. 
 The estimators and models of power spectrum as well as our fitting method have been tested using mock catalogues, and we find that they  perform well in recovering the fiducial values of the cosmological parameters of the simulations, and we also find that the errors of the parameters can be largely reduced by including the cross-power spectrum in the fit. We measure the auto-density, auto-momentum and cross power spectrum using the Sloan Digital Sky Survey Data Release 14  peculiar velocity catalogue.    The fit result of the growth rate $f\sigma_8$ is $f\sigma_8=0.413^{+0.050}_{-0.058}$ at effective redshift $z_{\mathrm{eff}}=0.073$, and our measurement is consistent  with the  prediction of the $\Lambda$ Cold Dark Matter cosmological model assuming General Relativity.  
The code of the power spectrum measurements,  power spectrum models  and window function convolution is in here: \url{https://github.com/FeiQin-cosmologist/PowerSpectrumMultipoles}. The Sloan Digital Sky Survey Data Release 14 peculiar velocity catalog (and mocks) used in this paper can be downloaded from: \url{https://zenodo.org/record/6640513}.
\end{abstract}


\section{Introduction}

Deciphering the formation and evolution of large-scale-structures in our Universe is paramount in cosmology. In recent years, several cosmological theoretical  models have been developed to understand the evolution of our Universe (\citealt{Peebles1980,Peebles1993,Peacock1999,Strauss1995} and references therein). The peculiar velocities of galaxies, arising from the fluctuation of the mass density field of the Universe, have been widely utilized to test these cosmological models. There are several substantial efforts that have been invested in this realm, including directly mapping the cosmic flow field, then comparing to the predictions from the cosmological models( e.g. \citealt{Nusser1995, Qin2019,Whitford2023} and references
therein), or alternatively, measuring the cosmological parameters and then comparing to the predictions of the cosmological models.  

The cosmological parameters that we seek to measure are the growth rate of
structure $f\equiv  d\ln\,D(a)/d\ln\,a$ which is the derivative of the linear growth factor $D(a)$ with respect to the  scale factor $a$, and the galaxy bias parameter $b\equiv  \sqrt{ P_{g}({\bf k})/P_{\mathrm{DM}}({\bf k})}$, which is the ratio between the galaxies density and dark matter density, where $P_{\mathrm{DM}}$ and $P_{g}$ are the dark matter and galaxies density power spectrum, respectively. Moreover,  the growth rate and galaxy bias parameter can be  normalized as $f\sigma_8$ and $b\sigma_8$, where $\sigma_8$ is defined as  the root mean square of the mass density fluctuation within spheres of 8 Mpc $h^{-1}$. Both $f\sigma_8$ and $b\sigma_8$ are measurable from galaxies.

To achieve this, there have been many techniques developed in  previous research.  One technique is estimating the growth rate $f\sigma_8$ and galaxy bias parameter $b\sigma_8$ from the galaxy and/or velocity correlation functions \citep{Gorski1989,Howlett2015,Adams2017,Dupuy2019,YuyuWang2018,YuyuWang2021,Turner2021,Turner2023}. However, this method faces two main challenges: the non-Gaussianity of velocity correlation, and the time-consuming nature of measuring the correlation function for large numbers of data-points or simulations. These parameters can also estimated via the comparison of measured/reconstructed velocity and density fields \citep{Pike2005,Erdogdu2006,Ma2012,Springob2014,Carrick2015,Said2020,Qin2023,Lilow2024}. This method has been shown to return smaller uncertainties on the growth rate, however, the systematic errors of the reconstruction methods are difficult to estimate and some recent work has demonstrated a potential for bias or underestimated error to exist in these previous measurements \citep{Turner2023b}.  A third "maximum likelihood fields" method, has been developed to estimate these parameters in recent years which also models the density and velocity fields directly, but remains computationally demanding.
 \citep{Adams2020,Lai2023,Carreres2023}. In particular \cite{Lai2023} used this technique to obtain constraints on the Sloan Digital Sky Survey Data Release 14 peculiar velocity catalogue (SDSSv, \citealt{Howlett2022}).

An alternative opportunity lies in measuring these parameters from the power spectrum of galaxies \citep{Feldman1994,Park2000,Park2006,Yamamoto2006,Blake2010,Johnson2014,Howlett2017,Zhang2017,Howlett2019,Qin2019b,Appleby2023,Shi2024}. This has the benefit of being faster to measure than the correlation function, and more natural to model from theory, but at the cost of introducing potential effects from complex survey geometries and shot-noise. The estimator for density power spectrum measurement was first developed in \cite{Feldman1994}, while the momentum power spectrum  was firstly investigated in \cite{Park2000} and \cite{Park2006}. Following these works, \cite{Howlett2019} (here after \papI) modified the momentum power spectrum method so that it can be implemented in the same way as the typical galaxy density power spectrum \citep{Yamamoto2006,Bianchi2015,Scoccimarro2015}. Further more, \cite{Qin2019b} (here after \papII) then applied this technique to successfully measure the cosmological parameters using the combined density and momentum power spectrum from the combined 2MASS Tully-Fisher (2MTF, \citealt{Hong2019}) and 6dFGS peculiar-velocity surveys (6dFGSv, \citealt{Campbell2014}). This paper is the third of the series. In this paper, we extend our method to measure and model the density and momentum \textit{cross} power spectrum, completing the full set of $3\times2$-point functions that can be measured from the two field. We use this to extract the cosmological parameters from the combined auto-density, auto-momentum and cross-power spectrum of the SDSSv catalogue, improving on the constraints from \cite{Lai2023}.   

This paper is structured as follows: in Section \ref{sec:datamock}, we introduce the SDSSv data and mock catalogues that reproduce the survey characteristics and are used to validate our analysis methods. In Section \ref{sec:psest}, we present the power spectrum estimators we apply to this data. In Section \ref{sec:psmod}, we introduce our power spectrum models and discuss how we account for the survey window function.  We test and verify our method in Section \ref{sec:psmock} on the mocks. Lastly, Section \ref{sec:psdata} presents our main results applying our analysis pipeline to the SDSSv data to estimate $f\sigma_8$ and $b\sigma_8$. We also discuss the consistency of these constraints with the predictions from GR and other works therein, before summarising in Section \ref{sec:conc}.  

Where necessary, we adopt a flat $\Lambda$ Cold Dark Matter ($\Lambda$CDM) fiducial cosmological model with parameters $n_s = 0.9653$, $\Omega_m=0.3121$,  $\Omega_b=0.0491$,  $\sigma_8=0.8150$ and   $H_0=100 h$ km s$^{-1}$ Mpc$^{-1}$, where $h=0.6751$. The corresponding fiducial value of the growth rate is $f\sigma_8=0.432$.

\section{The SDSS\lowercase{v} data and mocks}\label{sec:datamock}

\begin{figure}  
\includegraphics[width=\columnwidth]{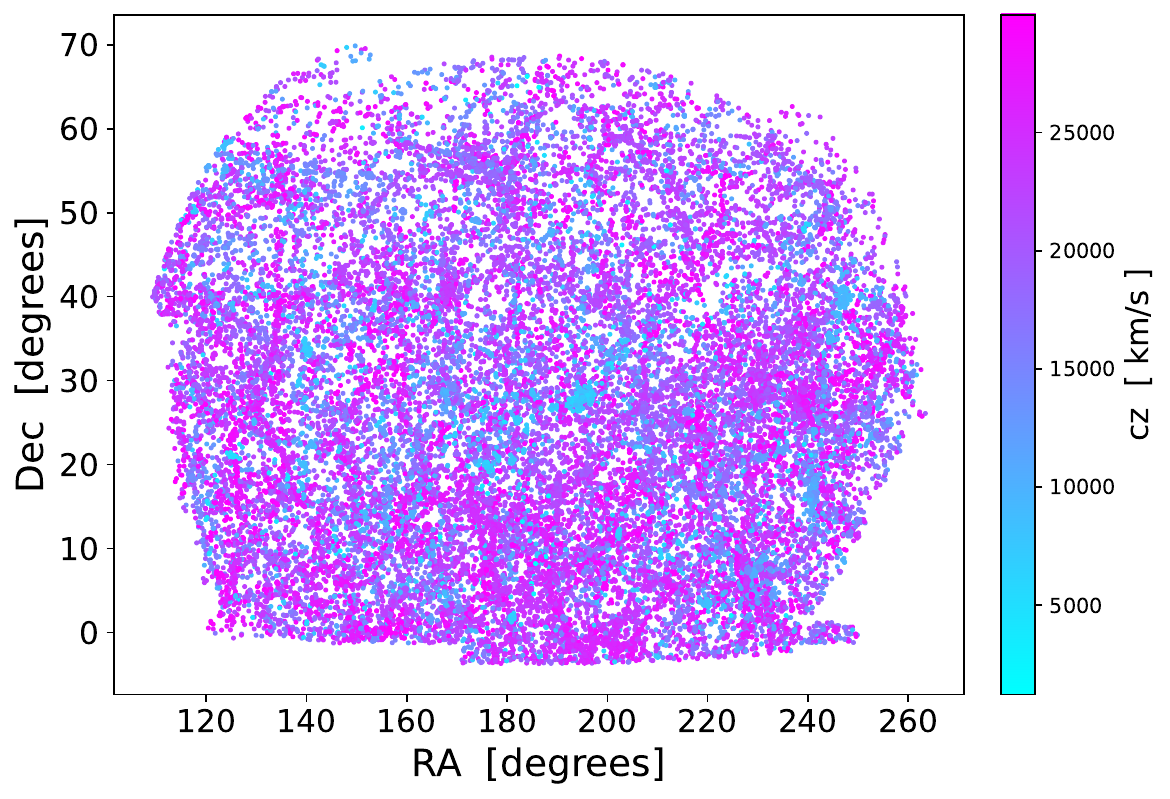}
 \caption{The sky coverage of the SDSSv galaxies. The color of the dot indicates the redshift of the galaxy according to the color bar.   }
 \label{wsky}
\end{figure}

\begin{figure}  
\includegraphics[width=\columnwidth]{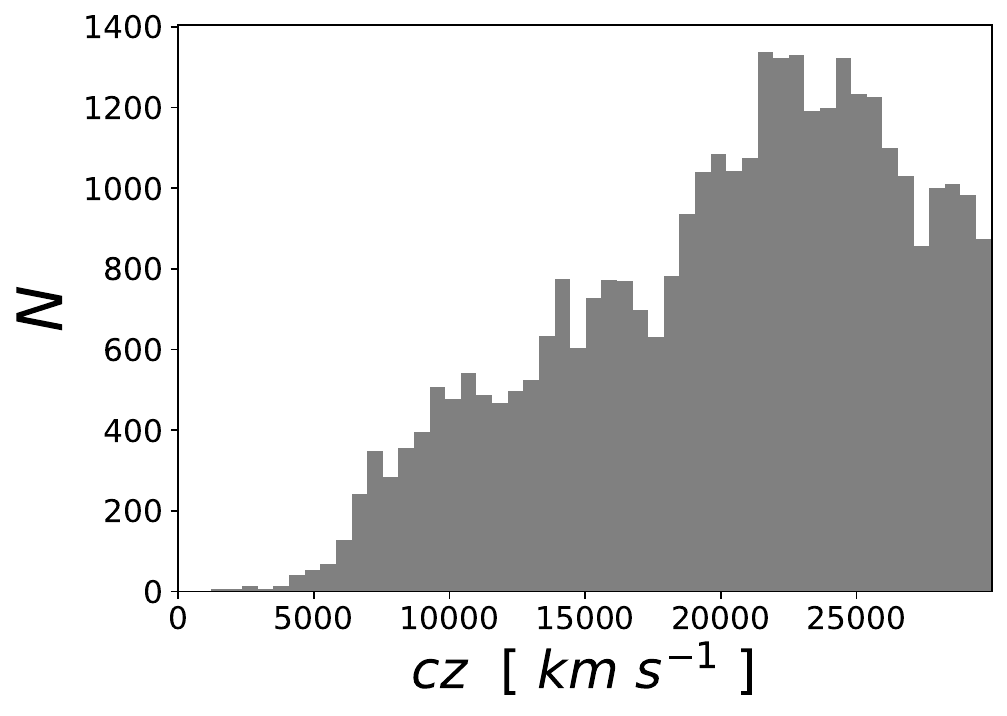}
 \caption{The redshift distribution of the SDSSv galaxies.  }
 \label{wcz}
\end{figure}

The Sloan Digital Sky Survey peculiar velocity catalogue (SDSSv, \citealt{Howlett2022})
 contains 34,059 early-type galaxies drawn from the larger SDSS Data Release 14 sample \citep{Abolfathi2018}. The sky coverage of the galaxies is shown in Fig.~\ref{wsky}. The Cosmic Microwave Background (CMB)-frame redshifts of the galaxies range from $z=0.0033$ to $z=0.1$, as displayed in Fig.~\ref{wcz}. The catalogue is compiled from galaxies with $r$-band apparent magnitudes  in the range  $10 < m_{r} \le 17$. In the catalogue, the logarithmic-distance ratio of a galaxy, which is defined as
\be\label{logd}
\eta \equiv \log_{10}\frac{d_z}{d_h}~,
\ee
 is estimated from the Fundamental Plane (FP, \citealt{Djorgovski1987}, \citealt{Dressler1987}, \citealt{Springob2014}) relation. $d_h$ is the true comoving distance of the galaxy inferred from position relative to the best-fit FP, while $d_z$ is the apparent distance of the galaxy obtained from the CMB-frame redshift.  In our paper, the line-of-sight peculiar velocities of galaxies are estimated from their $\eta$ using 
\citep{Johnson2014,Watkins2015, Adams2017, Howlett2017, Qin2018}
\be\label{watvp}
v=\frac{cz_{\mathrm{mod}} \ln10}{1+z_{\mathrm{mod}}}\eta,
\ee
where $z_{\mathrm{mod}}$ is given by \citep{Davis2014,Watkins2015}
\be  \label{zmod}
z_{\mathrm{mod}}=z\left[1+\frac{1}{2}(1-q_0)z-\frac{1}{6}(1-q_0-3q_0^2+1)z^2\right]\,,
\ee
and where the acceleration parameter is $q_0=0.5(\Omega_m-2\Omega_{\Lambda})=-0.532$. This relation uses the low-redshift approximation of the log-distance ratio and assumes that the true peculiar velocities of the galaxies are much less than their redshift.

\begin{figure}  
\includegraphics[width=\columnwidth]{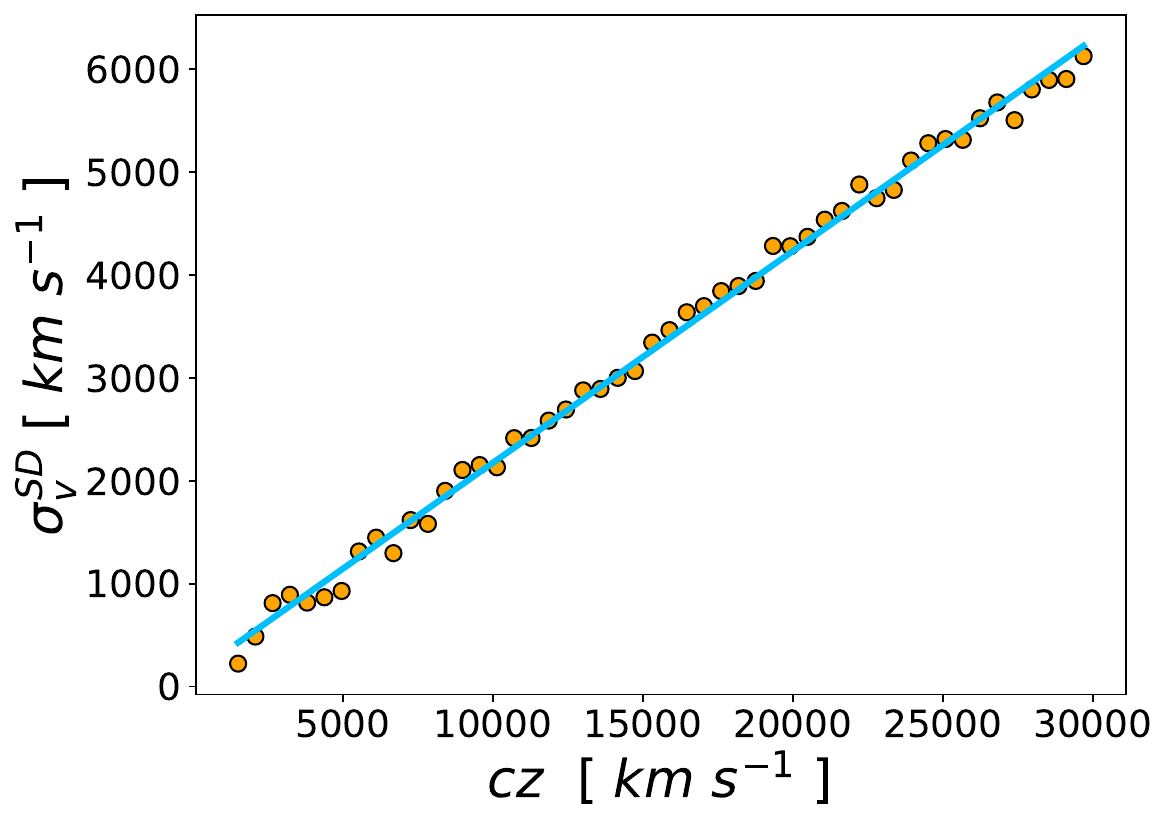}
 \caption{The standard deviation of peculiar velocities $\sigma_{v}^{SD}$ as a function of redshift $z$ for the SDSSv galaxies.   }
 \label{evrand}
\end{figure}

A substantial number of  mock galaxy catalogues is essential to verify the algorithm and estimate the errors of power spectrum and parameters. \cite{Howlett2022} provides 2048 mocks generated to match the real survey data. These mocks are created from the N-body dark
matter simulation code \textsc{l-picola} \citep{Howlett2015a,Howlett2015bs} using a halo occupation distribution model.  The mock sampling algorithm, presented in Section 3 of that paper, essentially builds upon the algorithms in \cite{Magoulas2012}, \cite{Scrimgeour2016} and \cite{Qin2018,Qin2019b,Qin2021a}.  These mocks can accurately reproduce the luminosity selection function, survey geometry, Fundamental Plane relation and galaxy clustering of the real SDSSv survey.

A random catalogue is also provided by \cite{Howlett2022}. To accurately extract the cosmological parameters from the power spectrum, the survey geometry of SDSSv must be taken into consideration properly --- measuring the power spectrum using fast-fourier transforms in the presence of a non-periodic and incomplete survey geometry results in a convolution of the true power spectrum with a survey window function. The random catalogue will be utilized to estimate the window function and apply the same convolution to our model during fitting of the data. Achieving this for the momentum and cross-power spectrum when weighting the data to obtain optimal constraints requires assigning fake velocities to the random points that match the velocity dispersion of the data. For that purpose, we firstly bin the real galaxies into 50 redshift bins, then calculate the standard deviation of the peculiar velocities in each redshift bin, as shown in the yellow-filled circles  in Fig.~\ref{evrand}. We fit a line to the measurements, as shown in the blue line in Fig.~\ref{evrand}, whose best-fit result is given by  
\be  \label{evR}
\sigma^{SD}_v=0.206~cz+117.182~.
\ee
Finally, in each redshift bin, we generate the velocities for the random points using a Gaussian function centered at zero and with standard deviation inferred from the above equation.  

  $\sigma_v^{SD}$ encapsulates the intrinsic scatter and measurement errors of the peculiar velocities. It is dominated by the measurement errors of peculiar velocities which can be very large values, usually on the order of 10${^3}$ km s$^{-1}$. On the other hand, $\sigma_v$ presented in Section \ref{sec:modps} is a velocity dispersion parameter which  accounts  for the non-linear motions of satellite galaxies in  their host halos as well as any other non-linear motions. This is typically assumed to be on the order of a few tens to hundreds km s$^{-1}$. However, the SDSSv galaxies are mainly early-type galaxies which often situated close to the centres of mass of their parent halos, and expected to have less non-linear motions. Therefore, it is not surprising the $\sigma_v$ we recover in this paper (see Table \ref{tabs22} or Fig.\ref{fsig8survey}) is found to be on the lower end of our usual expectations.   
   For the mock catalogs, $\sigma_v$ is actually not trivial to calculate and would require us to effectively re-run our algorithm for populating the dark matter halos with galaxies while storing additional parameters of this process. In our research, we treat $\sigma_v$ as a nuisance parameter in the fit, do find reasonable results compared to the expectation of 10-100 km/s and find our results on $f\sigma_8$ are insensitive to it's precise value. 
  
\section{The power spectrum estimation in redshift space}\label{sec:psest}
\subsection{The definition of field functions}\label{sec:psest1}

Following the method outlined in  \cite{Feldman1994}, we employ the following equation to estimate  the weighted galaxy density field
\be\label{fieldd}
F^{\delta}({\bf r})\equiv \frac{w_{\delta}({\bf r})\left[ n_{\delta}({\bf r})-\alpha n_s({\bf r}) \right]}{A_{\delta}} ~,
\ee
where  $w_p({\bf r})$ and $n_p({\bf r})$ are the weights and number of galaxies in the redshift sample at position ${\bf r}$.  In forming the density field we account for the survey
selection function by subtracting a random catalogue with  $n_s({\bf r})$ galaxies at ${\bf r}$ and where  $\alpha$ is the ratio of the total number of galaxies to
random points.

In addition, following the method outlined  in \papI, we estimate the weighed  galaxy momentum field using
\be\label{fieldp}
F^p({\bf r})\equiv \frac{w_p({\bf r})n_p({\bf r})v({\bf r})}{A_p} ~,
\ee
where $w_p({\bf r})$ and $n_p({\bf r})$ are the weights and number of galaxies in the peculiar velocity sample at location ${\bf r}$. $v({\bf r})$ is the measured line-of-sight peculiar velocity at this location.

For both of the fields the normalisation is defined  as 
\be 
A_{\delta,p}^2=\int w_{\delta,p}^2({\bf r})\bar{n}_{\delta,p}^2({\bf r}) d^3r~,
\ee 
where  $\bar{n}({\bf r})$ is the mean number density of galaxies at location ${\bf r}$. Compared to \papI~ and \papII, we here make the distinction between the weights, normalisation
and number density of galaxies used in defining the two fields as the galaxies within redshift and peculiar velocity samples may not be fully overlapping. Peculiar velocity measurements can only be obtained for particular classes of galaxies, and require much higher signal-to-noise observations to obtain compared to redshifts, and so typical surveys contain significantly fewer of these. When extracting the cross-power spectrum of the  density and momentum fields 
it is important to ensure the weights and normalisation account for this.\footnote{We note however, that in this work using SDSSv, we are in the special case where $\bar{n}_{\delta}=\bar{n}_{p}$.}

\subsection{The estimators of the power spectrum multipoles  }\label{sec:crsPSest123}

The Hubble recessional redshift and peculiar velocity of a galaxy contribute to form its observed redshift.  Therefore, the position of a galaxy inferred from its observed redshift is not its true comoving position. Instead, it is a `redshift-space' position contaminated by the peculiar motion. This is the so-called redshift-space distortion (RSD).   
The  spherical symmetry about the line of sight of the power spectrum   estimated from the galaxies' redshift-space positions is broken by RSD. Therefore, a common method is  to decompose  the redshift-space power spectrum using the Legendre polynomials $L_{\ell}(\mu)$, i.e. 
\be\label{plkest}
P^{\delta,p}({\bf k})= \sum_{\ell}P^{\delta,p}_{\ell}(k )L_{\ell}(\mu)
\ee
where ${\bf k}$ is the  wave vector.
The angular information of  ${\bf k}$ is encapsuled into $L_{\ell}(\mu)$, while the amplitude information of  ${\bf k}$ is enclosed into  $P^{\delta,p}_{\ell}(k )$ which are the so-called power spectrum multipoles. 
 The cosine of the angle between the     unit wave vector ${\bf\hat{k}}$ and the line-of-sight unit position vector ${\bf\hat{r}}$ is $\mu={\bf\hat{r}}\cdot {\bf\hat{k}}=\mathrm{cos}\,\theta$. The index `$\delta, p$' denotes the density and momentum power spectrum,  respectively. 

The estimators of the  density and  momentum power spectrum multipoles   have been derived in \cite{Yamamoto2006} and \papI~ respectively, given by
\be \label{autops}
P^{\delta,p}_{\ell}(k)=|F^{\delta,p}_{\ell}(k)|^2-N_{\ell}^{\delta,p}({\bf k})
\ee 
where the shot-noise terms are given by 
\be\label{Pnoised}
N^{\delta}_{\ell}= 
 (1+\alpha)\int w^2_{\delta}({\bf r})\bar{n}_{\delta}({\bf r})L_{\ell}(\mu) d^3r 
\ee
\be\label{Pnoisep}
N^{p}_{\ell}= \int w_{p}^2({\bf r})\bar{n}_p({\bf r})\langle v^2({\bf r}) \rangle L_{\ell}(\mu)d^3r 
\ee
respectively.  The shot-noise arises from the fact that galaxies are discrete tracers of the field. For the $\ell>0$ multipoles, the shot-noise is close to zero and can be ignored \citep{Yamamoto2006,Blake2010,Johnson2014,Howlett2019}.

The above mentioned density power spectrum is estimated from the density field only, while the momentum power spectrum is estimated from the momentum field only, therefore, they are ‘auto’ density power spectrum and ‘auto’ momentum power spectrum, respectively. 
In addition to the auto-power spectrum, we can also explore the cross-correlation between the density field and momentum field, which yields the conception of   the cross power spectrum of the density and momentum fields. 
 Given the definitions of the two weighted fields  we write the expectation value of the cross-correlation of the density and momentum fields
at different locations as 
\be  \label{eq4s}
\begin{split}
&\langle F^{\delta}({\bf r}) F^{p*}({\bf r}' ) \rangle\\
=& \frac{1}{ {A_{\delta}A_{p}}}  w_{\delta}({\bf r})w_{p}({\bf r}')\langle [ n_{\delta}({\bf r})-\alpha n_s({\bf r}) ] n_p({\bf r}' ) v({\bf r}' )  \rangle \\
=&\frac{1}{ {A_{\delta}A_{p}}} [ w_{\delta}({\bf r})w_{p}({\bf r}')\bar{n}_{\delta}({\bf r})\bar{n}_p({\bf r}')\xi_{\delta p}(|{\bf r}-{\bf r}'|) \\
& + w_{\delta}({\bf r})w_{p}({\bf r})\min\{\bar{n}_{\delta}({\bf r}),\bar{n}_p({\bf r})\} \langle v({\bf r}') \rangle
\delta^D(|{\bf r}-{\bf r}'|)  ]
\end{split}
\ee 
where `*' denotes the complex conjugate and $\xi_{\delta p}(|{\bf r}-{\bf r}'|)$  is the cross-correlation function between the density and momentum fields and $\delta^D(|{\bf r}-{\bf r}'|)$  is the Dirac delta function.  The second, shot-noise, term arises from the fact that the same galaxies may be included in both the estimation of the density field and the
momentum field and is proportional to whichever sample has fewer objects \citep{Smith2009}. A similar term can arise when estimating the cross-density power spectrum between two overlapping galaxy samples, however it is usually assumed that the two subsamples used in such an analysis are
mutually exclusive and so this shot-noise vanishes. This is not the case in the application of the density-momentum cross power in this paper.
One might also expect that the expectation value of the average velocity at any given location should be zero and so this vanishes. However,
this is rarely true for a real dataset given the presence of both a bulk flow and possible systematics.   In practice we can compute this directly
from the data using   $\langle v({\bf r})\rangle=1/N\sum_N v({\bf r})$.

From here we follow the procedure in \papI, defining the multipoles of the cross-power spectrum using 

\be\label{plkestcrs}
P^{\delta p}({\bf k})= \sum_{\ell}P^{\delta p}_{\ell}(k )L_{\ell}(\mu)~.
\ee 
Here, make careful distinction of the index `$\delta p$' from `$\delta,p$' of Eq.\ref{plkest},  `$\delta p$' denotes the density-momentum cross power spectrum while `$\delta,p$' denotes the auto-density and auto-momentum power spectrum respectively. 
Then, we can write our estimator under the `local plane-parallel approximation’ as (see Appendix \ref{sec:CRSest} for more details)
\be  
\begin{split}
    &|F^{\delta}(k)F^{p *}_{\ell}(k)|\\
    &=\frac{2\ell+1}{V}\int\frac{d \Omega_k}{4\pi}\int d^3r \int  d^3r' F^{\delta}({\bf r})F^{p}({\bf r}') L_{\ell}(\hat{{\bf k}}\cdot\hat{{\bf r}'})  e^{i {\bf k} \cdot ({\bf r}-{\bf r}') }.
\end{split}
\ee  

The expectation of this estimator  can be computed substituting Eq.~\ref{eq4s} and following the algebra in \papI or see Appendix \ref{sec:CRSest}
\be  \label{eq8s}
\begin{split}
    &\langle |F^{\delta}(k)F_{\ell}^{p *}(k)| \rangle=\frac{2\ell+1}{ {A_{\delta}A_{p}}}\times\\
    &\int\frac{d \Omega_k}{4\pi} \left[ \sum_{\ell'} \int\frac{d^3k'}{(2\pi)^3} P^{\delta p}_{\ell'}G^{\delta}({\bf k}- {\bf k}')S^{p*}_{\ell\ell'}({\bf k},{\bf k}')+N^{\delta p}_{\ell}( {\bf k} )\right] 
\end{split}
\ee  
where
\be  \label{GKKKKK}
G^{\delta}({\bf k}- {\bf k}')=\int w_{\delta}({\bf r})\bar{n}_{\delta}({\bf r})  e^{i {\bf r} \cdot ({\bf k}-{\bf k}') }  d^3r
\ee  
\be  \label{SPSSStt}
S^{p*}_{\ell\ell'}({\bf k},{\bf k}')=\frac{1}{V}\int w_{p}({\bf r})\bar{n}_{p}({\bf r}) L_{\ell}(\hat{{\bf k}}\cdot\hat{{\bf r}}) L_{\ell'}(\hat{{\bf k}}'\cdot\hat{{\bf r}})  e^{-i {\bf r} \cdot ({\bf k}-{\bf k}') }  d^3r
\ee 

 \be\label{Pnoisedp}
N^{\delta p}_{\ell}= \int w_{\delta}({\bf r})w_{p}({\bf r})min\{\bar{n}_{\delta}({\bf r})\bar{n}_{p}({\bf r})\}\langle v({\bf r}) \rangle L_{\ell}(\hat{{\bf k}}\cdot\hat{{\bf r}}) d^3r 
\ee
In the absence of any window function, Eq.~\ref{eq8s} reduces to $\langle |F^{\delta}(k)F_{\ell}^{p *}(k)| \rangle=
P^{\delta p}_{\ell}+N^{\delta p}_{\ell}( {\bf k} )
$. We can hence define our estimator of the
density-momentum cross power spectrum as  
\be  \label{crsps}
P^{\delta p}_{\ell}=\frac{1}{2}\mathrm{Im}\{|F^{p}(k)F_{\ell}^{\delta *}(k)|-|F^{\delta}(k)F_{\ell}^{p *}(k)| \}-N^{\delta p}_{\ell}( {\bf k} )
\ee 
which we have ensured is symmetric in our two fields.   Enforcing parity invariance in the cross-power spectrum shows that the density-momentum cross-power must be a purely imaginary-valued function.  Because the time-evolution of gravity and the initial conditions of the Universe remain unchanged
under the transformation $\delta \rightarrow -\delta$ and $v \rightarrow -v$,  the cross-power spectrum of the density and momentum fields must remain similarly
unchanged. This then leads to the identity  $P^{\delta p}_{\ell}=-P^{\delta p *}_{\ell}$ which is only true for a purely imaginary function.

\subsection{Fourier Transform of the field functions}

The Fourier Transform of the field function,  $F_{\ell}(k)$ in Eq.~\ref{autops} and \ref{crsps} has been well studied by \cite{Yamamoto2006} and \cite{Bianchi2015}. 
The full expressions of   
 the estimators for the power spectrum even-multipoles  $P_0({\bf k})$,  $P_2({\bf k})$ and $P_4({\bf k})$ are presented in Section 3 of \cite{Bianchi2015} and will not be repeated here
 . The full expressions of   
 the estimators for the power spectrum odd-multipoles  $P_1({\bf k})$ and   $P_3({\bf k})$ are  given by 
\be\label{P1k}
P_1(k)=\frac{3}{A^2}\int \frac{d\Omega_k}{4\pi}[F({\bf k}) F^*_1({\bf k}) ]  ,
\ee
\be\label{P3k}
P_3(k)=\frac{7}{2A^2}\int \frac{d\Omega_k}{4\pi}F({\bf k})\left[ 5F^*_3({\bf k})-3F^*_1({\bf k})  \right]  ,
\ee
where
\be\label{A0}
F({\bf k})=\int  F({\bf r}) e^{i {\bf k} \cdot {\bf r} }    d^3r,
\ee
\be  
F_{1}({\bf k})=\frac{1}{k}\left( k_xD_x+k_yD_y+k_zD_z \right)
\ee 
\be  
\begin{split}
F_{3}({\bf k})=\frac{1}{k^3} (& k^3_xE_{xxx}+k^3_yE_{yyy}+k^3_zE_{zzz}\\
+& 3k^2_xk_yE_{xxy}+3k^2_xk_zE_{xxz}\\
+& 3k^2_yk_xE_{yyx}+3k^2_yk_zE_{yyz}\\
+& 3k^2_zk_xE_{zzx}+3k^2_zk_yE_{zzy}+6k_xk_yk_zE_{xyz})
\end{split}
\ee 
and where
\be 
D_i=\int \frac{r_i}{r} F({\bf r}) e^{i {\bf k} \cdot {\bf r} }  d^3r~,~~(i=x,y,z)
\ee
\be  
E_{ijn}=\int \frac{r_ir_jr_n}{r^3} F({\bf r}) e^{i {\bf k} \cdot {\bf r} }  d^3r ~,~~(i,j,n=x,y,z)
\ee
In the absence of wide-angle or general relativistic effects only the even-multipoles of the auto-power spectrum are non-zero \citep{Blake2019,Beutler2019}. Under similar conditions, only the odd-multipoles of the cross power spectrum are non-zero. 

\subsection{FKP weight}\label{sec:fkp}

In the previous sections, the weight factor $w_{\delta,p}({\bf r})$ was left as a free parameter. However, following the discussion in 
\cite{Feldman1994} and \papI, 
by minimizing the fractional variance of the estimated power spectrum, one can obtain
the optimal weights for $w_{\delta,p}({\bf r})$ for the density and momentum fields,
\be\label{wopt}
w_{\delta}({\bf r},k)=\frac{1}{1+\bar{n}_{\delta}({\bf r})P^{\delta}_{FKP}(k)}, 
\ee
and
\be\label{woptp}
w_p({\bf r},k)= \frac{1}{\langle v^2({\bf r}) \rangle+\bar{n}({\bf r})P^{p}_{FKP}(k)},
\ee
respectively, where $\langle v^2({\bf r}) \rangle$ accounts for the typical measurement errors of the peculiar velocities.

For the SDSSv data, following the discussions  outlined  in \cite{Watkins2015} and \papI, we adopt 
\be\label{stdvs}
\langle v^2({\bf r}) \rangle=\left( \frac{\ln(10)cz}{1+z}\sigma_{\eta}(r)\right)^2+300^2\,\mathrm{km^{2}\,s^{-2}},
\ee
where $\sigma_{\eta}(r)$ is the measurement error of log-distance ratio $\eta$. $300^2$ km$^{2}$ s$^{-2}$ is the intrinsic scatter of the peculiar velocities arising from non-linearities, the choice of which does not affect our results significantly (please refer to the Appendix A of \papII~for more discussion). The $\langle v^2({\bf r}) \rangle$ of the random catalogue is calculated from 
$\langle v^2({\bf r}) \rangle=(\sigma_v^{SD})^{2}+300^2$ km$^2$ s$^{-2}$ where $\sigma^{SD}_v$ is calculated from Eq.\ref{evR}.

Following \papII~ and \cite{Turner2023}, 
 to achieve the optimum measurements we 
set  $P^{\delta}_{\mathrm{FKP}}=1600h^{-3}\,\mathrm{Mpc^{3}} $ and $P^{p}_{\mathrm{FKP}}=5\times 10^9h^{-3}\,\mathrm{Mpc^{3}\,km^{2}\,s^{-2}}$ for density and momentum fields, respectively. These are close to the expected values of the power spectra at the scales of interest for measuring the growth rate.

\section{THE THEORETICAL MODEL OF THE POWER
SPECTRUM}\label{sec:psmod}

\subsection{The models of power spectrum multipoles}\label{sec:modps}

The power spectrum multipoles are defined as integrals with respect to the Legendre polynomials:
 \be\label{psmod}
P_l(k)=(2\ell+1) \int^1_0P(k,\mu)L_l(\mu)d\mu~,
\ee 
which is basically the inverse Legendre transformation corresponding to  Eq.~\ref{plkest} (and \ref{plkestcrs}).

Following the discussion in \papI, 
the redshift-space power spectrum $ P(k,\mu)$ can be modelled using 
the non-linear perturbation theory in  \cite{McDonald2009}, \cite{Vlah2012,Vlah2013} and \cite{Okumura2014}.  
 The  model of the density power spectrum is  given by
\be \label{psmodd}
\begin{split}
P^{\delta}(k,\mu)&=P_{00} + \mu^2(2P_{01}+P_{02}+P_{11}) \\
& +\mu^4(P_{03}+P_{04}+P_{12}+P_{13}+\frac{1}{4}P_{22}) ,
\end{split} 
\ee 
In addition, the  model of momentum power spectrum is given by 
\be \label{psmodp}
P^{p}(k,\mu)=\frac{[a(z)H(z)]^2}{k^{2}}\big[P_{11}+\mu^2(2P_{12}+3P_{13}+P_{22})\big],
\ee 
While, the (imaginary) cross power spectrum model is given by
\be  \label{psmoddp}
\begin{split}
P^{\delta p}(k,\mu)&=\frac{a(z)H(z)}{k} \mu\Big[P_{01} + P_{02} + P_{11} \\
&+ \mu^2\big(\frac{3}{2}P_{03} + 2 P_{04} + \frac{3}{2} P_{12} + 2P_{13} + \frac{1}{2}P_{22}\big) \Big]~.  
\end{split}
\ee 
The $H(z)$ is the Hubble parameter at the effective redshift of the survey,  $a(z)$ is the scale factor at the effective redshift, they are converted from the cosmological parameters of the present-day Universe using the equations  in Section \ref{sec:resutdata}. 

The expressions of  $P_{mn}$, $(m,n=0,1,2,3,4)$ are lengthy and fully presented in the Appendix of \papI; we will not repeat  them here. However, there are three terms updated in this work which had small errors in their derivation or presentation in \papI, and they are presented as follows:
\be  
\begin{split}
P_{00} & = b_{1}^{2}D^{2}(P_{L} + 2D^{2}(I_{00} + 3k^{2}P_{L}J_{00})) + 2b_{1}D^{4}(b_{2}K_{00} + b_{s}K^{s}_{00}  ) \\
&+ D^{4}\left(\frac{1}{2}b_{2}^{2}K_{01} + \frac{1}{2}b^{2}_{s}K^{s}_{01} + b_{2}b_{s}K^{s}_{02}+2b_{3nl}\sigma^2_3P_L\right),
\end{split}
\ee
\be  
\begin{split}
P_{02} & = f^{2}b_{1}D^{4}(I_{02} + \mu^{2}I_{20} + 2k^{2}P_{L}(J_{02} + \mu^{2}J_{20}))\\
&- f^2k^{2}\sigma^{2}_{v}P_{00} + f^{2}D^{4}(b_{2}(K_{20}+\mu^{2}K_{30}) + b_{s}(K^{s}_{20}+\mu^{2}K^{s}_{30})),
\end{split}
\ee
\be  
\begin{split}
P_{12} & = f^{3}D^{4}(I_{12} + \mu^{2}I_{21} - b_{1}(I_{03} + \mu^{2}I_{30}) + 2k^{2}P_{L}(J_{02} + \mu^{2}J_{20})) \\
&- f^{2}k^{2}\sigma^{2}_{v}P_{01}+2f^3k^2D^4\sigma^2_v(I_{01}+I_{10}+3k^2P_{L}(J_{01}+J_{10})).  \\
\end{split}
\ee 
where $D(z)$ is the linear growth factor.\footnote{These equations supersede their definition in \papI, but differ only in higher order terms containing $\sigma^{2}_{v}$, $b_{s}$ or $b_{3,nl}$, which are treated as free parameters. We investigated using these expressions and those presented in \papI for our analysis of the SDSSv data, and found little difference in the cosmological constraints. As such, we do not expect these errors to have affected previous results.} Basically, $P_{mn}$ are the integrals over the linear power spectrum $P_{L}$ and contain the cosmological parameters that need to be fit. These parameters are the growth rate $f\sigma_{8}$, the linear  galaxy parameter  $b_{1}\sigma_{8}$ and non-linear galaxy bias parameters $b_{2}\sigma_{8}$ and $b_{3nl}\sigma_{8}$, as well as the velocity dispersion $\sigma_v$.\footnote{The non-linear galaxy bias $b_s$ is calculated from $b_1$ use the local Lagrangian relations, see Appendix of \papI. }  In this paper, the auto and cross-power spectrum is estimated from a single galaxy catalogue where each galaxy has a redshift and peculiar velocity measurement. Therefore, we do not make any distinction for the bias and velocity dispersion parameters  between the density and momentum fields. The \textsc{camb}\footnote{CAMB:~ \url{https://camb.info/}~.} package \citep{Lewis:1999bs} is employed  to generate  the linear power spectrum $P_L$ at redshift  $z = 0$  for the fiducial cosmology. 

\subsection{The window function convolution}

To accurately  extract the cosmological parameters from the power spectrum of galaxy surveys, the effects of survey geometry should be properly evaluated when fitting the power spectrum  models  to the measured power spectrum.   Therefore, the window function convolution should be applied to the power spectrum   models, given by  
\be 
{\bf P}^{c}(k)=  {\bf W} \cdot  {\bf P}(k')
\ee 
where we concatenate the models of power spectrum multipoles $ P_{\ell'}(k')$ $(\ell'=0,1,2,3,4,...)$ together to obtain a single model vector ${\bf P}(k')=[ P_{0}(k'),P_{1}(k'),P_{2}(k'),P_{3}(k'),P_{4}(k'),...]$. The $P^{c}(k)$ is the window function convolved power spectrum  multipole model,  and it is   the one that should be compared with the measured power spectrum multipoles.  
${\bf W}$ is the window function convolution matrix, whose the shape is $(N_{\ell}\times N_{k})\times(N_{\ell'}\times N_{k'})$. $N_k$ is the number of $k$-bins of each measured power spectrum multipole, while the $N_{k'}$ is the number of $k'$ values used to compute  each model power spectrum multipole $P_{\ell'}(k')$.  $N_{\ell}$ is the
number of measured power spectrum multipoles. $N_{\ell'}$ is the
number of model power spectrum multipoles. The off-diagonal blocks of ${\bf W}$ are not zeros,  the window function convolution of a given power spectrum  multipole is also affected by the other power spectrum  multipoles, which means it is important to model higher order multipoles than are being used in the fit. In this paper, we model up to $\ell=4$.

The window function convolution matrix
${\bf W}$ is calculated using the random catalogues.  
In this study, we employ the method oulined in \cite{Blake2018}. Each element of the matrix ${\bf W}  $ is given by
\be  \label{winmat}
{\bf W}_{\ell}(k, k')  =\frac{4\pi}{A^2}\int d\Omega_k\sum^{\ell}_{m=-\ell}Y^{\ell}_{m}(\hat{{\bf k}})\sum_{\ell'}\sum^{\ell'}_{m'=-\ell'}  \frac{I^{\ell\ell'}_{mm'}(\hat{{\bf k}},\hat{{\bf k}}')}{2\ell'+1}~, 
\ee 
where  
\be \label{Allmm}
I^{\ell\ell'}_{mm'}(\hat{{\bf k}},\hat{{\bf k}}')=\frac{V}{(2\pi)^3} \int Y^{\ell'\ast}_{m'}(\hat{{\bf k}}')\tilde{n}_w( {{\bf k}}- {{\bf k}}')\tilde{S}^{\ell\ell' \ast}_{mm'}( {\bf k}-{\bf k}') d^3k'~,
\ee 
\be \label{Smmll}
\tilde{S}^{\ell\ell' \ast}_{mm'}( {\bf k} )=\frac{1}{V}\int w({\bf r}) \bar{n}({\bf r})Y^{\ell}_{m}(\hat{{\bf r}})Y^{\ell' \ast}_{m'}(\hat{{\bf r}})e^{i{\bf k}\cdot{\bf r}} d^3r~,
\ee 
\be \label{nbfft}
\tilde{n}_w( {{\bf k}} )=\int w({\bf r}) \bar{n}({\bf r})  e^{i {\bf r} \cdot {\bf k}  }  d^3r~,
\ee 
and where $Y^{\ell}_{m}(\hat{{\bf r}})$ are the spherical harmonics,  $w({\bf r})$ is the FKP weight as introduced in Section \ref{sec:fkp} and $\bar{n}$ is  the expected number density of galaxies as presented in  Section \ref{sec:psest1}.

In the case of the auto-power spectrum, 
$w({\bf r})$ and $\bar{n}$ should be calculated using the equations of Section \ref{sec:psest} for the density and momentum fields, separately.  
But for   the cross-power spectrum, following the arguments in \cite{Blake2018}, the following replacement should be applied to the normalisation  factor of Eq.~\ref{winmat}
\be 
A^2 \rightarrow   A_pA_{\delta}
\ee 
in the meanwhile,  the following replacement should be applied to Eq.~\ref{Allmm}  
\be 
\begin{split}
&\tilde{n}_w(\hat{{\bf k}}-\hat{{\bf k}}')\tilde{S}^{\ell\ell' \ast}_{mm'}(\hat{{\bf k}}-\hat{{\bf k}}')
\rightarrow \\
 &\frac{1}{2}\left[\tilde{n}^{p}_w(\hat{{\bf k}}-\hat{{\bf k}}')\tilde{S}^{\delta, \ell\ell' \ast}_{mm'}(\hat{{\bf k}}-\hat{{\bf k}}')+\tilde{n}^{\delta}_w(\hat{{\bf k}}-\hat{{\bf k}}')\tilde{S}^{p,\ell\ell' \ast}_{mm'}(\hat{{\bf k}}-\hat{{\bf k}}') \right] 
 \end{split}
\ee 
where $\tilde{n}^{\delta}_w$ and $\tilde{n}^{p}_w$ are calculated using Eq.~\ref{nbfft} for the density and momentum fields respectively, while  $\tilde{S}^{\delta, \ell\ell' \ast}_{mm'}$ and $\tilde{S}^{p,\ell\ell'}_{mm'}$ are calculated using Eq.~\ref{Smmll} for the density and momentum fields, respectively.

\section{Tests on mock catalogues}\label{sec:psmock}

\subsection{Fitting method} 
In each $k$-bin, the probability distribution of the power spectrum measured from the 2048 mocks is usually not exactly Gaussian distributed. 
Particularly, \papI~ found significant non-Gaussianity in the distribution of simulated power spectrum measurements from in the momentum power spectrum for small $k$, which can bias the estimation of the cosmological parameters \citep{wang2018,Appleby2023}.
In the course of this work, we hence employ the Box-Cox transformation \citep{Box1964,Sakia1992}
to remove the non-Gaussianity of the power spectrum. The transformation is defined as
\be\label{boxc1}
T\equiv\left \{
\begin{aligned}
&\frac{(P+\Delta)^{\lambda}-1}{\lambda}, &\lambda \neq 0\\
&\ln(P+\Delta)~, &\lambda = 0
\end{aligned}
\right.
\ee
where $T$ is the Gaussianized power spectrum, $\Delta$ is a manually chosen small shift to the power spectrum $P$ in order to keep them being positive, which will not change the covariance \citep{Box1964, Qin2021b}. The value of the transformation parameter $\lambda$ for each $k$-bin is estimated by maximizing the following likelihood function \citep{Box1964}:
\be\label{boxclog}
L(\lambda) \sim (\lambda-1)\sum_i^N\mathrm{ln}(P_i)-\frac{N}{2}\mathrm{ln}\left( \frac{\sum_i(T_i(\nu)-\bar{T}(\nu))^2}{N}   \right),
\ee
where $N=2048$ is the number of mocks, $P_i$ is the measured power spectrum in a $k$-bin of the $i$-th mock, $T_i$ is the corresponding Gaussanized   power spectrum of the $i$-th mock for that $k$-bin and $\bar{T}(\nu)$ is the average of   $T_i$  in the $k$-bin of all mocks. The estimation method is clearly presented in Section 3 of \cite{Box1964} and Section 5 of \papI. 

As the covariance matrix of power spectrum is estimated from a finite number of mock surveys, this will add an uncertainty to the inverse of the covariance matrix. This is the so-called Hartlap effect \citep{Hartlap2007}. Following the argument in \cite{Sellentin2016}, instead of directly minimizing the $\chi^2$, the following revised $t$-distribution shall be applied as the likelihood function to estimate the parameters $\theta=[f\sigma_{8},b_{1}\sigma_{8},b_{2}\sigma_{8}, b_{3nl}\sigma_8, \sigma^2_v]$:      
\be\label{tdis}
\mathcal{L}(T|\theta) \propto |\boldsymbol{\mathsf{C}}|^{-\frac{1}{2}} \left[  1+\frac{ \chi^2(T|\theta)}{N-1}  \right]^{-\frac{N}{2}},
\ee
where the $\chi^2$ is given by $\chi^2(\boldsymbol{T}_{o}|\theta)=
\left[ {\bf T}_{o}-{\bf T}^{c}_{m}(\theta)\right]\boldsymbol{\mathsf{C}}^{-1}\left[{\bf T}_{o}-{\bf T}^{c}_{m}(\theta)\right]^T$,
and where the covariance matrix $\boldsymbol{\mathsf{C}}$ is calculated from the Gaussianized power spectrum. ${\bf T}^{c}_{m}$ is the Gaussianized and window function convolved model power spectrum, while ${\bf T}_{o}$ is the  
Gaussianized measured power spectrum. 
In this paper, the cosmological parameters are fitted from the combined auto-density power spectrum monopole,
auto-momentum power spectrum monopole and cross-power spectrum dipole, i.e. ${\bf T}_{o}=[T^{\delta}_0,T^{p}_0,T^{\delta p}_1]$ which is a $3N_k$ vector,  ${\bf T}^c_{m}=[T^{\delta~c}_0,T^{p~c}_0,T^{\delta p~c}_1]$, which is a $3N_k$ vector too and $\boldsymbol{\mathsf{C}}$ is a $3N_k\times 3N_k$ matrix.   The other measured  power spectrum multipoles are close to zeros and noisy, therefore, we do not use them to constrain the cosmological  parameters in this paper.

The Metropolis-Hastings Markov-chain Monte Carlo (MCMC) algorithm is implemented to estimate the cosmological parameters.
We use flat priors in the interval $f\sigma_8\in(0,1.5]$, $b_1\sigma_8\in[0 ,3]$, $b_{2}\sigma_{8}\in[-3,3]$,  $b_{3nl}\sigma_{8}\in[-3,3]$ and $\sigma^2_{v}\in(0,250^2]$ km$^2$ s$^{-2}$ .

\begin{table*}   \centering
\caption{  The best-fit estimated values of the cosmological parameters from the SDSSv mocks.  These are fits to the average of 2048 mocks.  The number of degrees of freedom (d.o.f) is calculated based on the 28 data points for each power spectrum and 5 free parameters.  }
\begin{tabular}{|c|c|c|c|c|c|c|}
\hline
\hline
  &$f\sigma_8$  &$b_1\sigma_8$&$b_2\sigma_8$&$b_{3nl}\sigma_8$&$ \sigma^2_v$ $[km^{2}~s^{-2}]$  &$\chi^2/$d.o.f \\
\hline
$P^{p}_0$      & $0.382^{+0.105}_{-0.092}$& $1.697^{+0.883}_{-0.880}$& $<2.565$ &$<2.767$ &  $<48.1$  & $9.445/23$ \\ & & & & & & 
\\
$P^{\delta}_0$+$P^{p}_0$   & $0.422^{+0.087}_{-0.109}$& $1.141^{+0.080}_{-0.094}$& $-0.610^{+0.270}_{-0.241}$ 
&  $0.327^{+0.387}_{-0.453}$ 
& $62.7^{+37.3 }_{-24.0 }$ & 
$2.741 /51$ 
\\ & & & & & &
\\
$P^{\delta}_0$+$P^{p}_0$+$P^{\delta p}_1$  & $0.441^{+0.041}_{-0.038}$& $1.163^{+0.076}_{-0.065}$ & $-0.587^{+0.137}_{-0.124}$ & $0.264^{+0.213}_{-0.233}$& $82.0^{+14.3}_{-11.2}$&  $4.348/79$  \\[1ex]
\hline
\end{tabular}
 \label{tabs}
\end{table*}

\subsection{Tests on mock catalogues}\label{sec:mockstt}

\begin{figure*}  
\includegraphics[width=86mm]{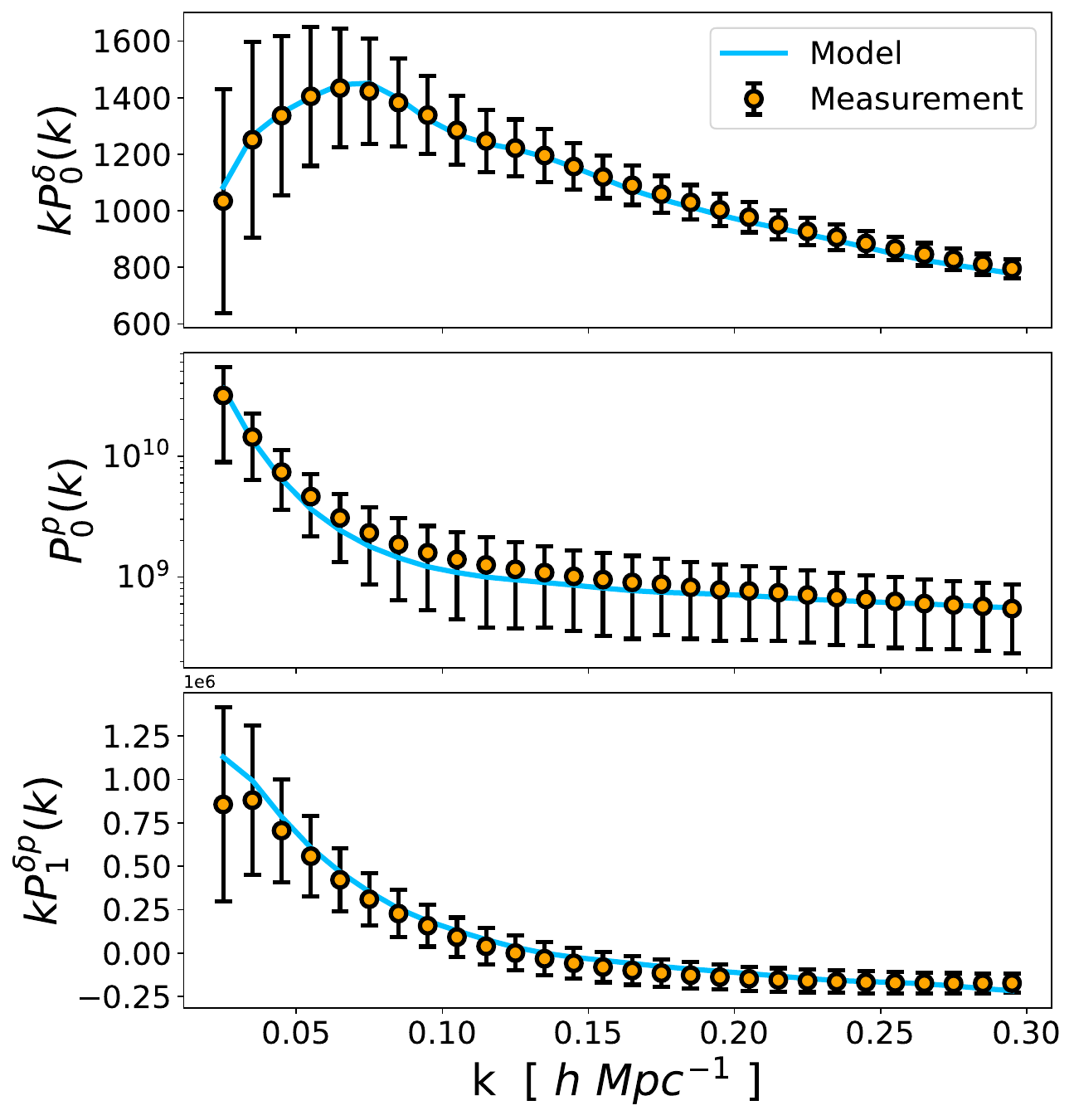}
\includegraphics[width=95mm]{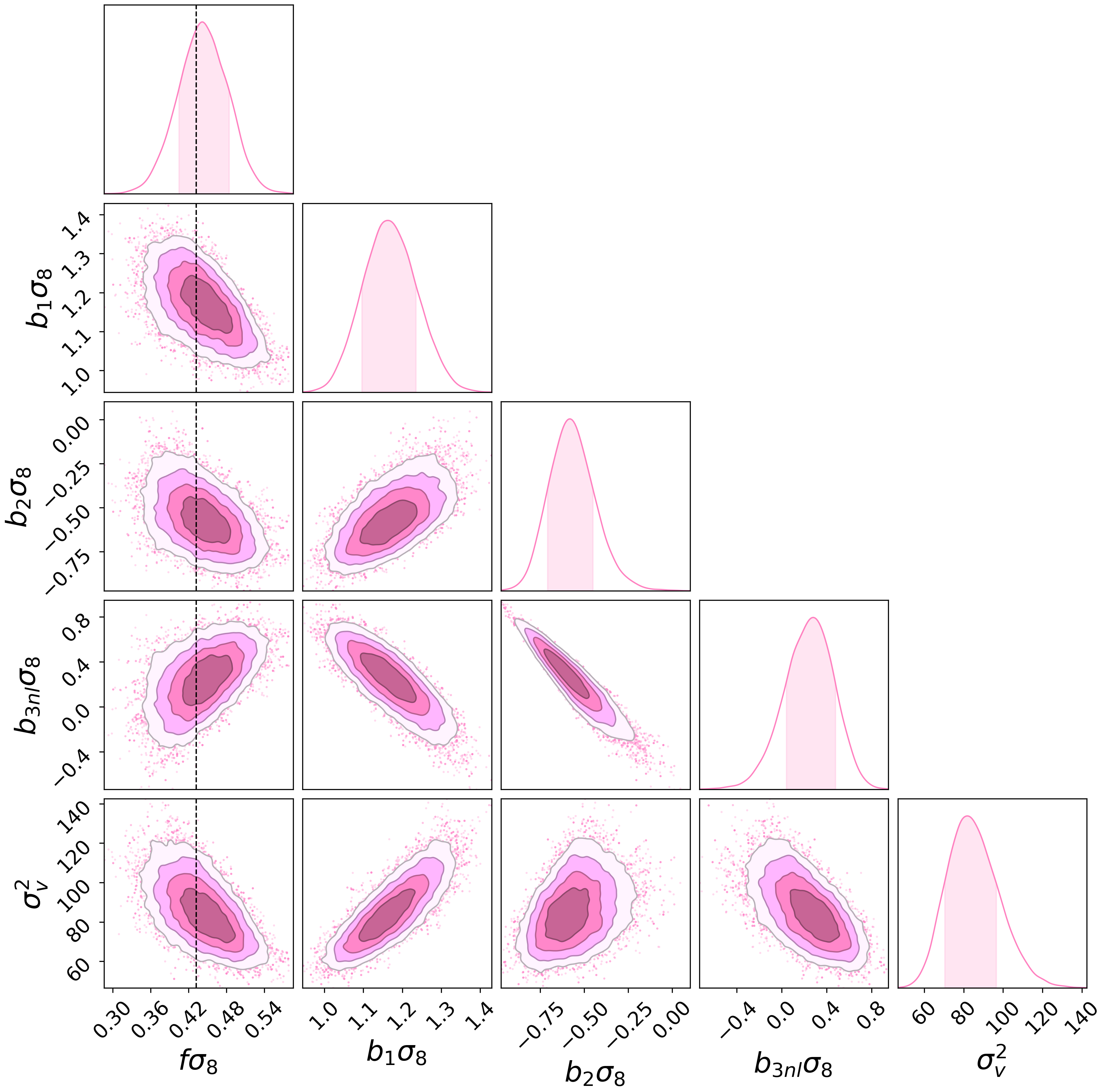}
 \caption{The power spectrum and parameter fit results of
the SDSSv mocks. In the left panels, the yellow filled circles in the top, middle and bottom panels are the measured density monopole, momentum monopole and cross dipole power spectrum, respectively. These are the average of the measurements of the 2048 mocks, with error bars representative of the error on a single realisation. The blue curves are the model power spectrum fit to the measurements. The right panel shows the
marginalised histograms and 2D contours of the MCMC samples of the cosmological parameters, the fit
results of the cosmological parameters are shown in  the top of the histograms.  The 2D contours indicate the 1, 1.5, 2 and 2.5$\sigma$ regions, whilst the shaded region in the histograms is $1\sigma$ confidence level.  The vertical dashed line indicates the fiducial value $f\sigma_8$=0.432. The corresponding marginalised values of the parameters are presented in the last row of Table \ref{tabs}.  }
 \label{fsig8mock}
\end{figure*}

To test the model and  the fitting methods, firstly, we measure the growth rate $f\sigma_8$     using the SDSSv mocks.
Fig.~\ref{fsig8mock} shows the fit results of the power spectrum of the SDSSv mocks. In the left-hand panels, the yellow filled circles in the top, middle and bottom panels are the measured density monopole, momentum monopole and cross dipole power spectrum, respectively. They are the average of the measurements of the 2048 mocks, where the uncertainties represent the expected error on a single realisation. The blue curves are the model power spectrum fit to the measurements.   The fit results of the cosmological parameters are shown in the last row of Table \ref{tabs}. The recovered value of the growth rate is $f\sigma_{8}=0.441^{+0.041}_{-0.038}$ which agrees well with the truth value (vertical dashed line). This indicates that our method can successfully recover the fiducial growth rate of the mocks and the model works well. 
 
In Table \ref{tabs}, we present the best fit values of the 
$f\sigma_{8}$, $b_{1}\sigma_{8}$, $b_{2}\sigma_{8}$, $b_{3nl}\sigma_8$ and $\sigma^2_v$ for different fitting configurations. If only considering the momentum power spectrum monopole $P^{p}_0$, the constraints on the two biasing parameters $b_{2}\sigma_{8}$ and $b_{3nl}\sigma_{8}$ and velocity dispersion $\sigma^2_v$ are weak compared to 
the fit results using the combined density and momentum power spectrum monopoles $P^{\delta}_0$+$P^{p}_0$ as well as the combined density monopole, momentum monopole and cross dipole  power spectrum $P^{\delta}_0$+$P^{p}_0$+$P^{\delta p}_1$.  We find that the momentum power spectrum alone can however constrain $f\sigma_{8}$ reasonably well, with the density power spectrum serving primarily to break the large-scale degeneracy with the bias parameters. Nonetheless, the combination of all the three power spectrum yields the tightest fit results for the parameters, and the error of $f\sigma_8$ can be reduced by $\sim 55\%$  when including the cross-power spectrum. 

For the calculation of the $\chi^2$ in Table \ref{tabs}, we 
use the mean over 2048 catalogs as the data vector, with the errors estimated for 1 realization.  However, scaling up the $\chi^2$ by 2048 gives a value that is 
too large. This is simply because while the model is clearly accurate enough for our current data volume, it is not accurate enough at the level of 2048 times of our volume. If one were to attempt to fit real data with such a large cosmological volume, they would indeed find a poor fit. Fortunately, we are not yet at the level where this is a systematic we need to consider. 

\begin{figure}  
\includegraphics[width=\columnwidth]{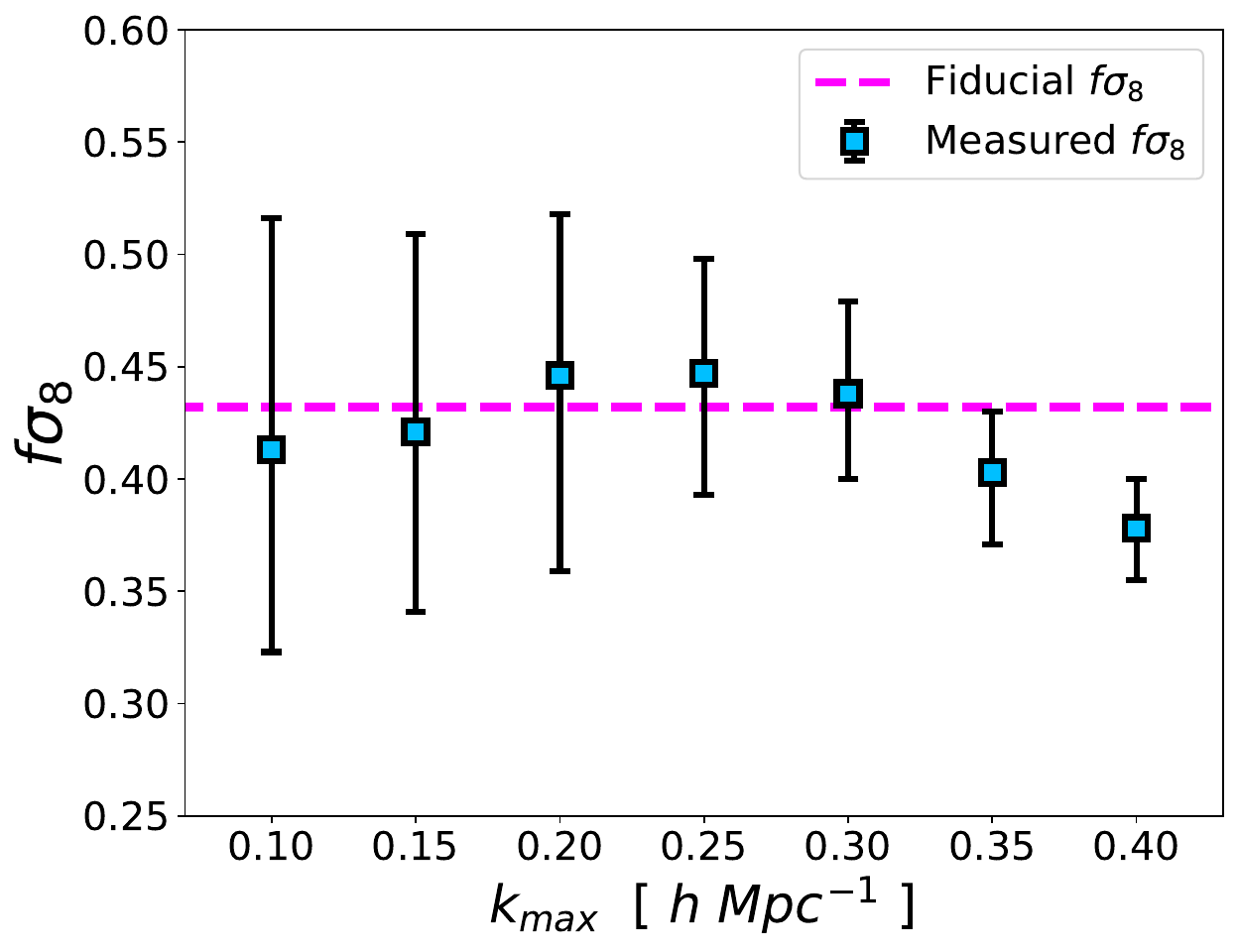}
 \caption{The blue filled-squares show the estimated growth rate $f\sigma_8$ as a function of cut-off wave number $k_{\mathrm{max}}$. The pink dashed-line is the fiducial value $f\sigma_8$=0.432. In all cases, the minimum fitting scale is set to $k=0.025h~\mathrm{Mpc}^{-1}$.}
 \label{fsig8vsk}
\end{figure}

Our growth rate constraints are obtained from measuring and fitting the power spectra in bins of 
$k\in[0.025,~k_{\mathrm{max}}]$ with bin widths of $0.01 h\,\mathrm{Mpc^{-1}}$.
 In Fig.~\ref{fsig8vsk}, the blue filled squares  showcase the recovered $f\sigma_8$ as a function of the cut-off $k_{\mathrm{max}}$ we used. The pink dashed line indicates  the fiducial value  $f\sigma_8=0.432$. As indicated by Fig.~\ref{fsig8vsk},
 a $k_{\mathrm{max}}$ value around 0.3$h~\mathrm{Mpc}^{-1}$ is the best choice for the fit since it gives both smaller systematic error and smaller measurement error.  If $k_{\mathrm{max}}\geq$0.35$h~\mathrm{Mpc}^{-1}$, the  measurement error on $f\sigma_8$ is significantly reduced, but the  systematic offset is  significantly increased due to the power spectrum model failing to accurately reproduce these scales.  For smaller $k_{\mathrm{max}}\leq$0.15$h~\mathrm{Mpc}^{-1}$, the measurement error on $f\sigma_8$ is large due to a large fraction of the information, especially on the higher order parameters $b_2\sigma_8$ and $b_{3nl}\sigma_8$ and their correlation with $f\sigma_8$, existing on non-linear scales. Therefore,  $k_{\mathrm{max}}$=0.3$h~\mathrm{Mpc}^{-1}$ is adopted for the fit of the growth rate to the real data in our paper.

\section{ RESULTS AND discussion}\label{sec:psdata}

\subsection{Results}\label{sec:resutdata}
\begin{figure*}  
\includegraphics[width=86mm]{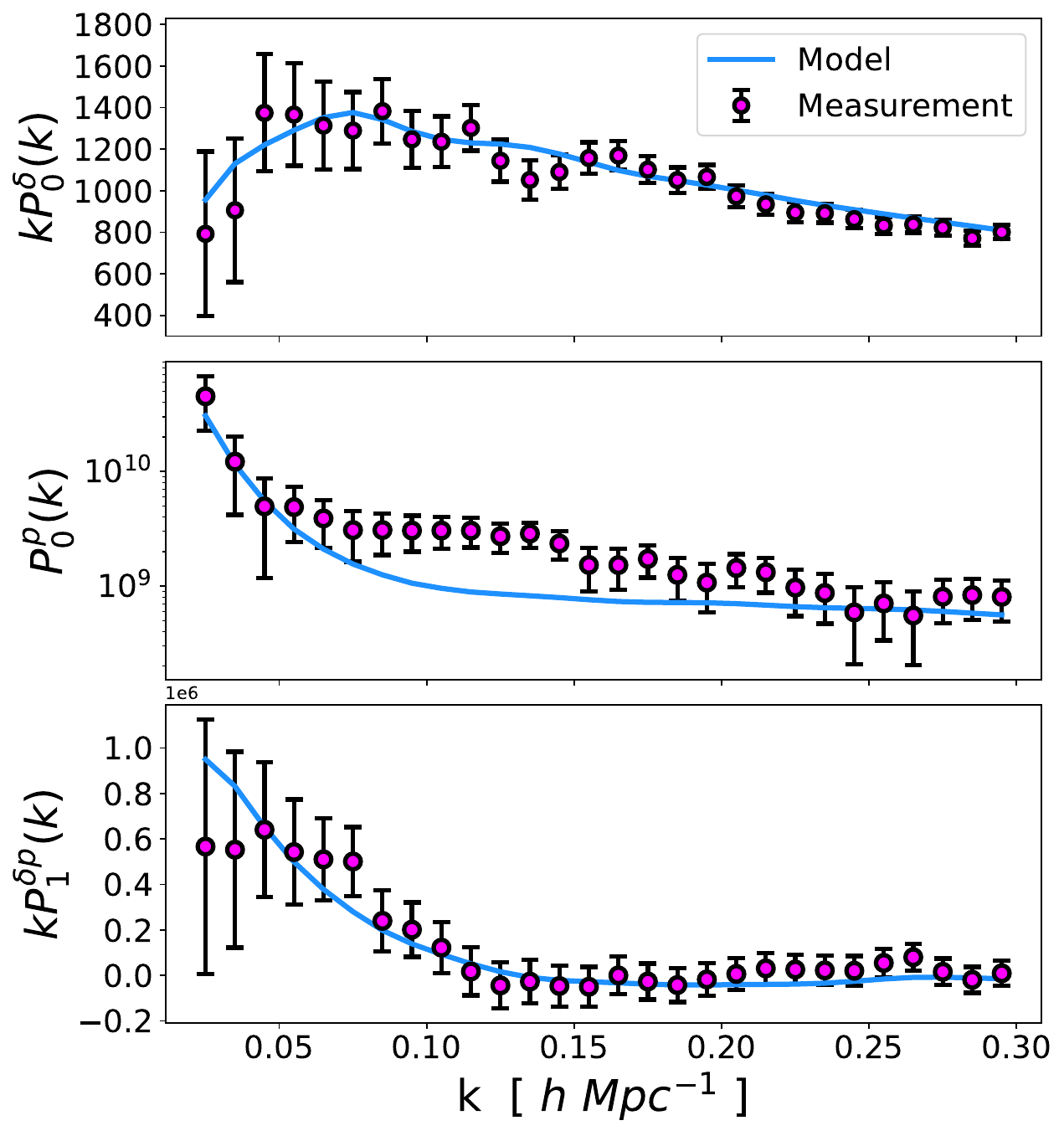}
\includegraphics[width=95mm]{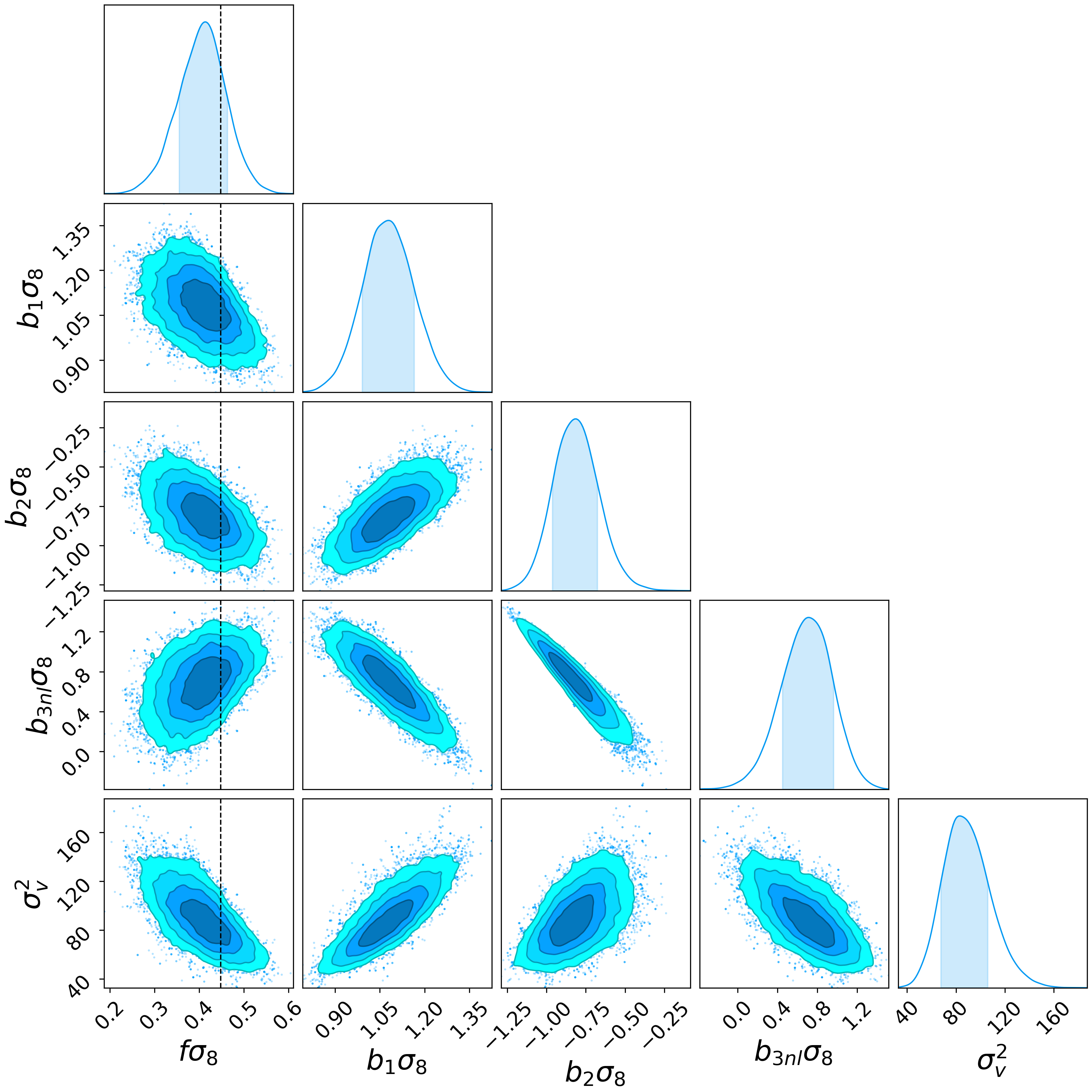}
 \caption{Same as Fig.~\ref{fsig8mock} but for the SDSSv survey data. The vertical dashed line indicates the GR+\cite{Planck2018} prediction, $f\sigma_8=0.448$. The corresponding best-fit estimated values of the parameters are presented in Table \ref{tabs22}.}
 \label{fsig8survey}
\end{figure*}

Fig.~\ref{fsig8survey} shows the fit results of the power spectra of the SDSSv data.  In the left-hand-side panels, the pink filled circles in the top, middle and bottom panels are the measured density monopole, momentum monopole and cross dipole power spectra, respectively.  The blue curves are the model power spectrum fit to the measurements. The corresponding $\chi^2/\mathrm{d.o.f}=72.894/(84-5)=0.923$.  The fit results of the cosmological parameters are shown in   Table \ref{tabs22}. The best-fit estimated value of the growth rate is
$f\sigma_8=0.413^{+0.050}_{-0.058}$.

Assuming General Relativity (GR), 
 following the method outlined in  \cite{Linder2007} and 
\cite{Howlett2015},
the growth  rate at redshift $z$ can be calculated using
\be  \label{fsig8th}
f(a)\sigma_8(a)=\Omega_{m}(a)^{\gamma}\sigma_{8}\frac{D_{gr}(a*)}{D_{gr,0}}\frac{D_{\gamma}(a)}{D_{\gamma}(a*)},
\ee
where
\be  
\frac{D_{\gamma}(a)}{D_{\gamma}(a*)}=\mathrm{exp}\left( \int^a_{a*}\Omega_{m}(a')^{\gamma}dlna' \right).
\ee
\be 
D_{gr}(a)=\frac{H(a)}{H_0}\int^a_0\frac{da'}{a'^3H(a')^3}, 
\ee 
\be 
\frac{H(a)}{H_0}=E(a)\equiv\sqrt{\frac{\Omega_{m0}}{a^3}+\frac{1-\Omega_{m0}-\Omega_{\Lambda0}}{a^2}+\Omega_{\Lambda0}}
\ee 
and where
\be \label{ahz}
a=\frac{1}{1+z}~,~~\Omega_{m}(a)=\frac{\Omega_{m0}}{a^3E(a)^2}.
\ee 
$\gamma=0.55$ assuming GR, and remaining parameters of the cosmological model are the Hubble constant $H_0$, present-day matter density parameter $\Omega_{m0}$ and present-day dark energy density parameter $\Omega_{\Lambda0}$. Assuming best-fit $\Lambda$CDM \cite{Planck2018}   values for the cosmological parameters\footnote{We use the chain \url{base_plikHM_TTTEEE_lowTEB_lensing_1.txt} on \url{https://irsa.ipac.caltech.edu/data/Planck/release_2/ancillary-data/HFI_Products.html} as a prior
on $\Omega_{m}(z)$.}, one can obtain the GR predicted growth rate at our effective redshift $z_{\mathrm{eff}}=0.073$ of $f\sigma_8=0.448$, which is consistent with  our measurement $f\sigma_8=0.413^{+0.050}_{-0.058}$ at the 68\% confidence level.

\begin{table}   \centering
\caption{  The best estimated values of the cosmological parameters of the SDSSv data.    }
\begin{tabular}{|c|c|c|   }
\hline
\hline
  $f\sigma_8$  &$0.413^{+0.050}_{-0.058}$\\
\hline
 $b_1\sigma_8$ & $1.078^{+0.086}_{-0.087}$ \\ 
 \hline
 $b_2\sigma_8$ & $-0.810^{+0.141}_{-0.150}$\\ 
 \hline
 $b_{3nl}\sigma_8$ & $0.711^{+0.261}_{-0.260}$\\ 
 \hline
 $ \sigma^2_v$ $[km^{2}~s^{-2}]$& $82.0^{+24.0}_{-14.1}$ \\ 
 \hline
 $\chi^2/$d.o.f & $72.894/79$ \\ 
\hline
\end{tabular}
 \label{tabs22}
\end{table}


\subsection{Discussion}

\begin{figure*}  
\includegraphics[width=180mm]{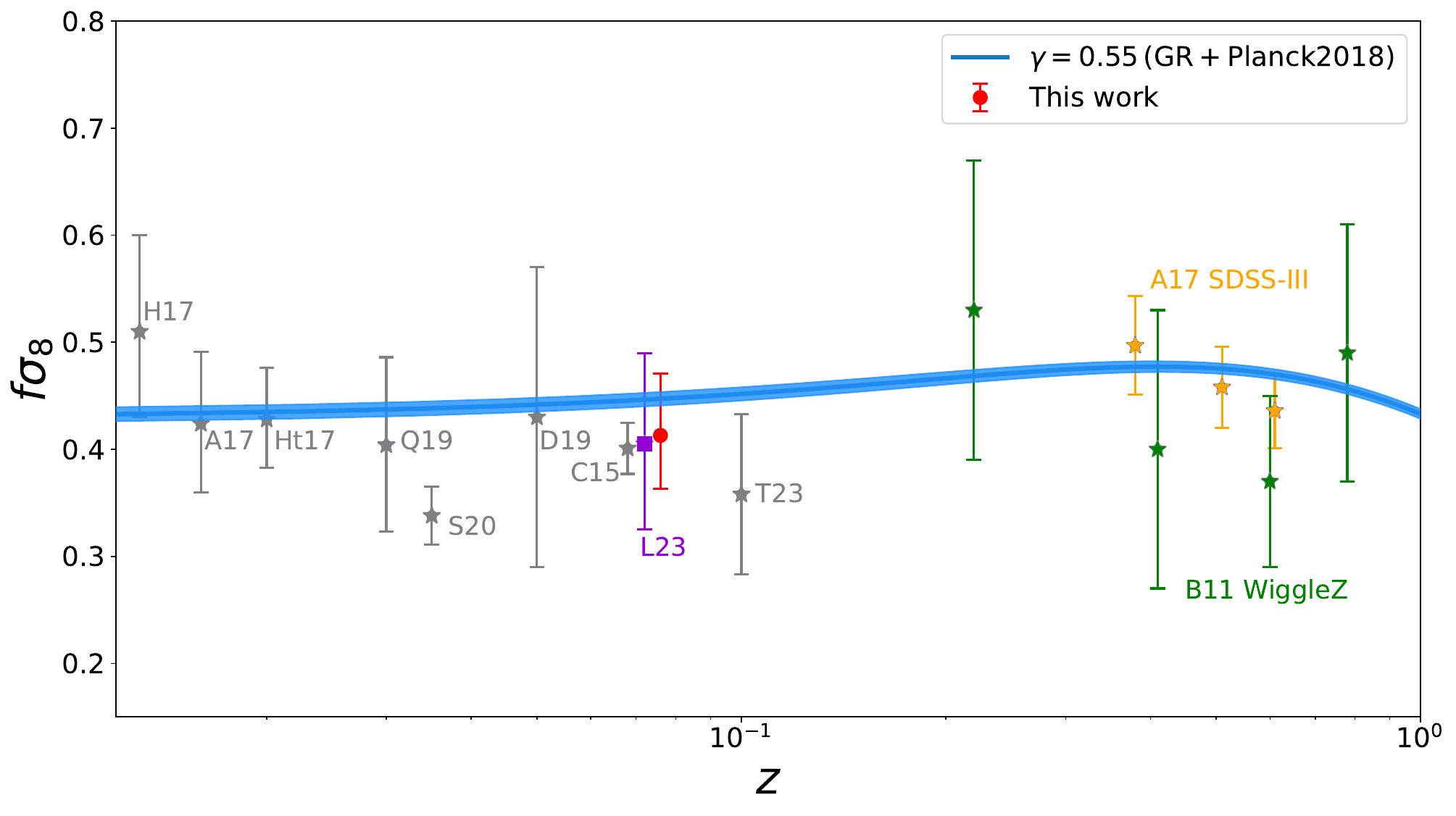}
 \caption{
 The growth rate $f\sigma_8$ as a function of redshift $z$. The  blue curve is computed using the equations \ref{fsig8th} to \ref{ahz} assuming GR and a \cite{Planck2018}-based $\Lambda$CDM cosmology. The red dot is the measurement of this paper. The purple filled square (L23) is \citet{Lai2023} using the same  data as this paper. The gray, yellow and green pentagrams show the measurements from other surveys: 
H17:\citet{Howlett2017} using 2MTF;
A17:\citet{Adams2017} using 6dFGS;
Ht17: \citet{Huterer2017}  using JLA+CSP and SN+6dFGSv;
Q19:\papI~ using combined 2MTF and 6dFGSv; 
S20:\citet{Said2020} using 6dFGS+SDSS;
D19:\citet{Dupuy2019} uisng Cosmicflows-3;
C15:\citet{Carrick2015} using 2M++, SFI++ and A1;
T23:\citet{Turner2023} using 6dFGS; 
B11 WiggleZ:\citet{Blake2011} using WiggleZ (four green stars);
A17 SDSS-III: \citet{Alam2017} using SDSS-III (three yellow stars).  
}
 \label{fsig8vZ}
\end{figure*}

Fig.~\ref{fsig8vZ} shows measurements and theory predictions for $f\sigma_8$ as a function of the redshift $z$. Following the method in \papII, the blue curve is computed using the equations of Section \ref{sec:resutdata}. We again employ the \cite{Planck2018} cosmological model. The gray, yellow and green pentagrams show the measurements from other surveys \citep{Blake2011, Carrick2015, Adams2017, Alam2017, Howlett2017, Huterer2017, Dupuy2019, Qin2019b, Said2020, Turner2023}. 
 Although each measurement is mostly consistent with the \cite{Planck2018}-based prediction, when taken as an ensemble there are hints that the measurements are lower than the prediction. This could be resolved with a slightly lower $\Omega_m$ value, or a larger $\gamma$ \citep{Macaulay2013,Said2020,Quelle2020}.

\cite{Lai2023}, the purple filled (L23) square in Fig.~\ref{fsig8vZ},  used the same survey data as this paper to measure the growth rate using the maximum likelihood fields method, reporting a measured value of $f\sigma_8=0.405^{+0.085}_{-0.080}$ at the same effective redshift. This has a slightly larger error than our measurement, since in their method they only using nonlinear models of the power spectrum for $k_{max}$=0.15$h~Mpc^{-1}$ to extract information on the growth rate. As discussed in Section \ref{sec:mockstt}, $k_\mathrm{max}$=0.15$h~Mpc^{-1}$ is not large enough to include all the available growth rate information and as a result, their measurement error is larger. For $k_\mathrm{max}$=0.15$h~Mpc^{-1}$, the    growth rate fitted from our power spectrum is $f\sigma_8=0.442^{+0.085}_{-0.091}$ , which is very comparable to \cite{Lai2023} in 68\% confidence level.

\section{Conclusions}\label{sec:conc}

In this paper, we constrain the cosmological parameters using the power spectrum measured from the redshifts and peculiar velocities data in the Sloan Digital Sky Survey Data Release 14 peculiar velocity catalogue. 
We extend the previous work of \papI~ and \papII, and develop a method to measure and model the cross power spectrum of the density and momentum fields, which has only odd multipoles and is complementary to the auto power spectra of the two fields.

We have tested the full set of $3\times 2$-pt power spectrum estimators, models and parameter fitting methods on a set of simulations that reproduce the characteristics of the SDSSv data. We find these methods perform well in recovering the fiducial value of the growth rate of the mocks, up to a $k_\mathrm{max}=0.3h~Mpc^{-1}$. We also find that including the cross-power spectrum improves the constraints by $\sim55\%$ compared to fitting just the two auto-power spectra.
 
Applying our validated method to the SDSSv data, the growth rate fitted from the combined  density, momentum and the cross power spectrum is
 $f\sigma_8=0.413^{+0.050}_{-0.058}$ at redshift $z_{\mathrm{eff}}=0.073$. This measurement is consistent  with the prediction of General Relativity for a \cite{Planck2018}-based $\Lambda$CDM cosmological model and with previous results in the same redshift range using different techniques and/or data. 

The code of the power spectrum measurements,  power spectrum models  and window function convolution can be found in here:
\url{https://github.com/FeiQin-cosmologist/PowerSpectrumMultipoles}

The Sloan Digital Sky Survey Data Release 14 peculiar velocity catalogue and mocks used in this paper can be downloaded from:
\url{https://zenodo.org/record/6640513}

\acknowledgments
The project leading to this publication has received funding from Excellence Initiative of Aix-Marseille University - A*MIDEX, a French ``Investissements d'Avenir'' program (AMX-20-CE-02 - DARKUNI).
CH is supported by the Discovery Project (projects FL180100168 and DP20220101395)  
of the Australian Research Council’s Laureate Fellowship.

%



\

\software{  
      \textsc{ChainConsumer} \citep{ChainConsumer},
           \textsc{emcee} \citep{Foreman-Mackey2013}, 
          \textsc{Scipy} \citep{Virtanen2020}, 
          \textsc{Matplotlib} \citep{Hunter2007},
           \textsc{CAMB} \citep{Lewis:1999bs}.   
        }

\

\

\appendix
 
\section{The Galilean transformation of  $P^{p}$ and $P^{\delta p}$} \label{sec:bh}

In our paper series, the density, momentum and cross-power spectrum are originally defined as
\be  
(2\pi)^3\delta^D({\bf k}-{\bf k}')P^{\delta}({\bf k})\equiv\langle  \delta({\bf k}) \delta({\bf k}') \rangle
\ee 
\be \label{defmomps}
(2\pi)^3\delta^D({\bf k}-{\bf k}')P^{p}({\bf k})\equiv\langle (1+\delta({\bf k}))v({\bf k})(1+\delta({\bf k}'))v({\bf k}')\rangle
\ee 
\be
(2\pi)^3\delta^D({\bf k}-{\bf k}')P^{\delta p}({\bf k})\equiv\langle  (1+\delta({\bf k}))v({\bf k})  \delta({\bf k}') \rangle
\ee 
respectively and where $\delta^D({\bf k}-{\bf k}')$ is the Dirac $\delta$-function.  The momentum power spectrum $P^{p}({\bf k})$ and cross-power spectrum $P^{\delta p}({\bf k})$ are not invariant under the Galilean transformation, i.e, they
depend on the bulk motion of the sample with
respect to the frame\footnote{In our research, the peculiar velocities of the mock galaxies and real galaxies are calculated based on the CMB frame.}.  

In particular, for the momentum power spectrum, if we shift the line-of-sight peculiar velocities of the sample by a constant bulk velocity $\epsilon$, i.e.
$v'({\bf k})=v({\bf k})+\epsilon$, Eq.~\ref{defmomps} gives
 \be 
 \begin{split} 
&(2\pi)^3\delta^D({\bf k}-{\bf k}')P^{p}_G({\bf k})\\
\equiv&\langle (1+\delta({\bf k}))v'({\bf k})(1+\delta({\bf k}'))v'({\bf k}')\rangle\\
=&\langle (1+\delta({\bf k}))(v({\bf k})+\epsilon)(1+\delta({\bf k}'))(v({\bf k}')+\epsilon)\rangle \\
=&\langle 
(1+\delta({\bf k}))v({\bf k})(1+\delta({\bf k}'))v({\bf k}')\rangle +\epsilon^2\langle\delta({\bf k})\delta({\bf k}')\rangle+2\epsilon\langle (1+\delta({\bf k}))v({\bf k})\delta({\bf k}'))\rangle + \epsilon\langle (1+\delta({\bf k}))v({\bf k})\rangle 
 +\epsilon^2\langle  1+2\delta({\bf k}') \rangle\\
=&(2\pi)^3\delta^D({\bf k}-{\bf k}')\left[ P^{p}({\bf k})+\epsilon^2P^{\delta}({\bf k})+2\epsilon P^{\delta p}({\bf k}) \right]+\epsilon^2   + \epsilon\langle (1+\delta({\bf k}))v({\bf k})\rangle 
 +2\epsilon^2\langle\delta({\bf k}) \rangle\\
 \end{split}
\ee  
where the average of the density contrast field $\langle\delta({\bf k}) \rangle $ and the average of the momentum field $\langle (1+\delta({\bf k}))v({\bf k})\rangle$   are assumed to be zero. Then the above equation is reduced to
\be \label{Galipp}
P^{p}_G({\bf k})=P^{p}({\bf k})+\epsilon^2P^{\delta}({\bf k})+2\epsilon P^{\delta p}({\bf k})+\epsilon^2
\ee  
Therefore, under the Galilean transformation, the momentum power spectrum of Eq.~\ref{defmomps} will be transformed to Eq.~\ref{Galipp}.
Similarly, under the Galilean transformation, the cross-power spectrum will be transformed to
\be \label{Galipdp}
P_G^{\delta p}({\bf k})= P^{\delta p}({\bf k})+\epsilon P^{\delta}({\bf k})  
\ee 
which is also not invariant under the Galilean transformation.

The bulk velocity $\epsilon$ encapsulates: (1) any possible bulk motions of the galaxies caused by the gravitational fluctuations; and (2) any possible systematic errors introduced from the observations and measurements.  
Therefore,  we will explore whether our fit results are sensitive to the bulk velocity $\epsilon$.
Plugging the expressions of power spectrum models (presented in Section \ref{sec:psmod}) into 
Eq.~\ref{Galipp} and \ref{Galipdp}, yields the  Galilean transformed    power spectrum models. Then we fit these models 
to the measured power spectrum of the SDSSv data. We set $\epsilon$
as a free parameter and use flat prior in the interval $\epsilon\in[-1000,1000]$ km s$^{-1}$.

\begin{figure}  
\includegraphics[width=90mm]{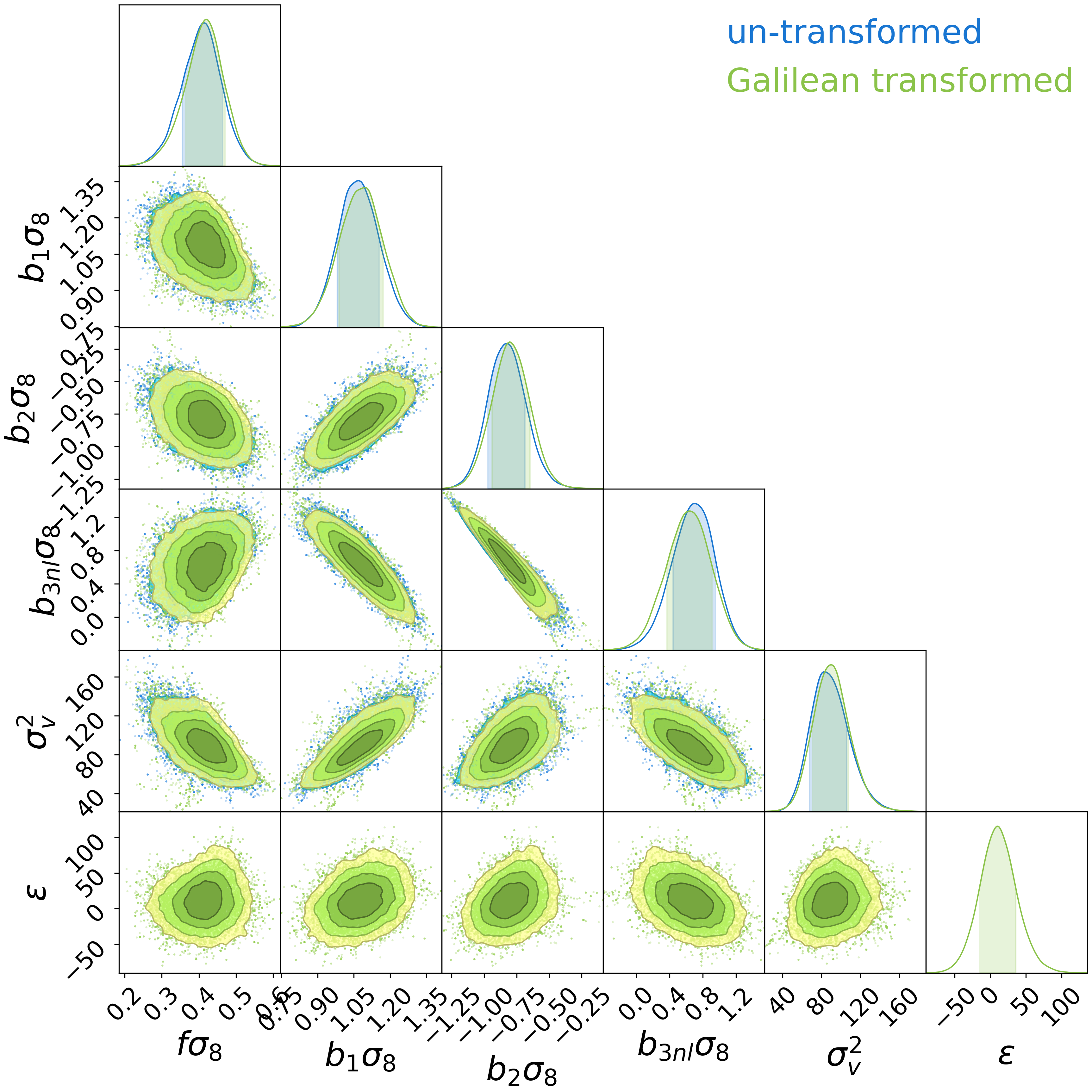}
 \caption{ The green-colored histograms and 2D-contours show the fit results of the parameters by adding   $\epsilon$.  The blue-colored histograms and 2D-contours  show the fit results without $\epsilon$ (same as Fig.\ref{fsig8survey}).}
 \label{Galilo}
\end{figure}

The fit results are shown in the green-colored histograms and 2D-contours of Fig.\ref{Galilo}. For comparison,  the blue-colored histograms and 2D-contours displays  the fit results without Galilean transformation (same as Fig.\ref{fsig8survey}).
The fit results with and without Galilean transformation are consistent, and the best estimated value of $\epsilon$ is $16^{+25}_{-26}$ km s$^{-1}$ which is consistent with zero. 
Therefore,  the systematic shift $\epsilon$ is negligible.

\section{The estimator of the cross-power spectrum}\label{sec:CRSest}

Corresponding to Section \ref{sec:crsPSest123}, we can write our estimator of the cross-power spectrum under the `local plane-parallel approximation’ as
\be  
\begin{split}
    |F^{\delta}(k)F^{p *}_{\ell}(k)|=&\int F^{\delta}({\bf r})e^{i {\bf k} \cdot{\bf r}} d^3r \int  \Big[\frac{2\ell+1}{V}\int F^{p*}({\bf r}')L_{\ell}(\hat{{\bf k}}\cdot\hat{{\bf r}'})\frac{d \Omega_k}{4\pi}\Big]e^{-i {\bf k} \cdot{\bf r}'}d^3r'\\
    =&\frac{2\ell+1}{V}\int\frac{d \Omega_k}{4\pi}\int d^3r \int  d^3r' F^{\delta}({\bf r})F^{p*}({\bf r}') L_{\ell}(\hat{{\bf k}}\cdot\hat{{\bf r}'})  e^{i {\bf k} \cdot ({\bf r}-{\bf r}') }.
\end{split}
\ee  
Plugging Eq.\ref{eq4s} into the above equation to replace $F^{\delta}({\bf r})F^{p*}({\bf r}')$, one can obtain
\be  
\begin{split}
    \langle|F^{\delta}(k)F^{p *}_{\ell}(k)|\rangle 
    =&\frac{1}{ {A_{\delta}A_{p}}} \frac{2\ell+1}{V}\int\frac{d \Omega_k}{4\pi}\int d^3r \Big[\int  d^3r'  w_{\delta}({\bf r})w_{p}({\bf r}')\bar{n}_{\delta}({\bf r})\bar{n}_p({\bf r}')\xi_{\delta p}(|{\bf r}-{\bf r}'|) L_{\ell}(\hat{{\bf k}}\cdot\hat{{\bf r}'})  e^{i {\bf k} \cdot ({\bf r}-{\bf r}') }\\
& +  \int  d^3r'w_{\delta}({\bf r})w_{p}({\bf r})\min\{\bar{n}_{\delta}({\bf r}),\bar{n}_p({\bf r})\} \langle v({\bf r}') \rangle
\delta^D(|{\bf r}-{\bf r}'|)  ]L_{\ell}(\hat{{\bf k}}\cdot\hat{{\bf r}'})  e^{i {\bf k} \cdot ({\bf r}-{\bf r}') }\Big]
\end{split}
\ee  
The second term of the above equation integrals over the Dirac delta function yields
\be  \label{FFest}
\begin{split}
    \langle|F^{\delta}(k)F^{p *}_{\ell}(k)|\rangle =&\frac{1}{ {A_{\delta}A_{p}}} \frac{2\ell+1}{V}\int\frac{d \Omega_k}{4\pi}\int d^3r \Big[\int  d^3r'  w_{\delta}({\bf r})w_{p}({\bf r}')\bar{n}_{\delta}({\bf r})\bar{n}_p({\bf r}')\xi_{\delta p}(|{\bf r}-{\bf r}'|) L_{\ell}(\hat{{\bf k}}\cdot\hat{{\bf r}'})  e^{i {\bf k} \cdot ({\bf r}-{\bf r}') }\\
& +   w_{\delta}({\bf r})w_{p}({\bf r})\min\{\bar{n}_{\delta}({\bf r}),\bar{n}_p({\bf r})\} \langle v({\bf r}) \rangle
 L_{\ell}(\hat{{\bf k}}\cdot\hat{{\bf r}})   \Big]
\end{split}
\ee 
If we define the shot-noise term as
\be 
N^{\delta p}_{\ell} = \int   w_{\delta}({\bf r})w_{p}({\bf r})\min\{\bar{n}_{\delta}({\bf r}),\bar{n}_p({\bf r})\} \langle v({\bf r}) \rangle
 L_{\ell}(\hat{{\bf k}}\cdot\hat{{\bf r}})  d^3r
\ee 
which is Eq.\ref{Pnoisedp}, 
Eq.\ref{FFest} can be reduced to
\be \label{FFxi}  
\begin{split}
    \langle|F^{\delta}(k)F^{p *}_{\ell}(k)|\rangle =&\frac{1}{ {A_{\delta}A_{p}}} \frac{2\ell+1}{V}\int\frac{d \Omega_k}{4\pi}\Big[\int d^3r \int  d^3r'  w_{\delta}({\bf r})w_{p}({\bf r}')\bar{n}_{\delta}({\bf r})\bar{n}_p({\bf r}')\xi_{\delta p}(|{\bf r}-{\bf r}'|) L_{\ell}(\hat{{\bf k}}\cdot\hat{{\bf r}'})  e^{i {\bf k} \cdot ({\bf r}-{\bf r}') } +   N^{\delta p}_{\ell}    \Big]
\end{split}
\ee 
The two point correlation is related to power spectrum through
\be 
\xi_{\delta p}(|{\bf r}-{\bf r}'|)=\frac{1}{(2\pi)^3}\int P^{\delta p}({\bf k})e^{-i {\bf k} \cdot ({\bf r}-{\bf r}') } d^3k
\ee
Plugging the above equation into Eq.\ref{FFxi} to replace $\xi_{\delta p}$, we obtain
\be  \label{iugh87}
\begin{split}
    &\langle|F^{\delta}(k)F^{p *}_{\ell}(k)|\rangle \\
    =&\frac{1}{ {A_{\delta}A_{p}}} \frac{2\ell+1}{V}\int\frac{d \Omega_k}{4\pi}\Big[\int d^3r \int  d^3r'  w_{\delta}({\bf r})w_{p}({\bf r}')\bar{n}_{\delta}({\bf r})\bar{n}_p({\bf r}') \Big( \int P^{\delta p}({\bf k}')e^{-i {\bf k}' \cdot ({\bf r}-{\bf r}') } \frac{d^3k'}{(2\pi)^3} \Big) L_{\ell}(\hat{{\bf k}}\cdot\hat{{\bf r}'})  e^{i {\bf k} \cdot ({\bf r}-{\bf r}') }  + N^{\delta p}_{\ell} \Big]\\
    =&\frac{1}{ {A_{\delta}A_{p}}} \frac{2\ell+1}{V}\int\frac{d \Omega_k}{4\pi}\Big[\int d^3r \int  d^3r'  w_{\delta}({\bf r})w_{p}({\bf r}')\bar{n}_{\delta}({\bf r})\bar{n}_p({\bf r}')   \int \frac{d^3k'}{(2\pi)^3} P^{\delta p}({\bf k}')     L_{\ell}(\hat{{\bf k}}\cdot\hat{{\bf r}'})  e^{i    ({\bf k}-{\bf k}')\cdot {\bf r} } e^{-i   ({\bf k}-{\bf k}')\cdot{\bf r}' } + N^{\delta p}_{\ell}\Big]\\
    =&\frac{1}{ {A_{\delta}A_{p}}} \frac{2\ell+1}{V}\int\frac{d \Omega_k}{4\pi}\Big[\Big(\int   w_{\delta}({\bf r})\bar{n}_{\delta}({\bf r}) e^{i    ({\bf k}-{\bf k}')\cdot {\bf r} } d^3r \Big)\int  d^3r' w_{p}({\bf r}')\bar{n}_p({\bf r}')   \int \frac{d^3k'}{(2\pi)^3} P^{\delta p}({\bf k}')     L_{\ell}(\hat{{\bf k}}\cdot\hat{{\bf r}'})  e^{-i   ({\bf k}-{\bf k}')\cdot{\bf r}' } + N^{\delta p}_{\ell}\Big]\\
\end{split}
\ee  
If define the following function
\be 
G^{\delta}({\bf k}-{\bf k}')\equiv \int   w_{\delta}({\bf r})\bar{n}_{\delta}({\bf r}) e^{i    ({\bf k}-{\bf k}')\cdot {\bf r} } d^3r
\ee
which is Eq.\ref{GKKKKK}, then 
Eq.\ref{iugh87} can be reduced to
\be  
\begin{split}
    \langle|F^{\delta}(k)F^{p *}_{\ell}(k)|\rangle=&\frac{1}{ {A_{\delta}A_{p}}} \frac{2\ell+1}{V}\int\frac{d \Omega_k}{4\pi}  \Big[ \int  d^3r'   \int \frac{d^3k'}{(2\pi)^3} P^{\delta p}({\bf k}')     L_{\ell}(\hat{{\bf k}}\cdot\hat{{\bf r}'})  e^{-i   ({\bf k}-{\bf k}')\cdot{\bf r}' } w_{p}({\bf r}')\bar{n}_p({\bf r}') G^{\delta}({\bf k}-{\bf k}')+ N^{\delta p}_{\ell} \Big]\\
\end{split}
\ee  
Expanding the power spectrum $P^{\delta p}({\bf k}')$ of the above using Eq.\ref{plkestcrs}, then the above yields
\be  \label{r232gg}
\begin{split}
    &\langle|F^{\delta}(k)F^{p *}_{\ell}(k)|\rangle\\
    =&\frac{1}{ {A_{\delta}A_{p}}} \frac{2\ell+1}{V}\int\frac{d \Omega_k}{4\pi}  \Big[ \int  d^3r'   \int \frac{d^3k'}{(2\pi)^3} \Big(\sum_{\ell'}P^{\delta p}_{\ell'}(k' )L_{\ell'}(\hat{{\bf k}}'\cdot\hat{{\bf r}'})\Big)    L_{\ell}(\hat{{\bf k}}\cdot\hat{{\bf r}'})  e^{-i   ({\bf k}-{\bf k}')\cdot{\bf r}' } w_{p}({\bf r}')\bar{n}_p({\bf r}') G^{\delta}({\bf k}-{\bf k}')+ N^{\delta p}_{\ell} \Big]\\
    =&\frac{2\ell+1}{ {A_{\delta}A_{p}}} \int\frac{d \Omega_k}{4\pi}  \Big[    \sum_{\ell'} \int \frac{d^3k'}{(2\pi)^3}   P^{\delta p}_{\ell'}(k' ) G^{\delta}({\bf k}-{\bf k}') \frac{1}{V}\int  d^3r' w_{p}({\bf r}')\bar{n}_p({\bf r}') L_{\ell'}(\hat{{\bf k}}'\cdot\hat{{\bf r}'}) L_{\ell}(\hat{{\bf k}}\cdot\hat{{\bf r}'})  e^{-i   ({\bf k}-{\bf k}')\cdot{\bf r}' } + N^{\delta p}_{\ell} \Big] \\
\end{split}
\ee  
If we define the following function
\be 
S^{p*}_{\ell\ell'}({\bf k},{\bf k}')=\frac{1}{V}\int   w_{p}({\bf r})\bar{n}_p({\bf r})  L_{\ell}(\hat{{\bf k}}\cdot\hat{{\bf r} })L_{\ell'}(\hat{{\bf k}}'\cdot\hat{{\bf r} })  e^{-i   ({\bf k}-{\bf k}')\cdot{\bf r}  } d^3r
\ee 
which is Eq.\ref{SPSSStt}, the Eq.\ref{r232gg} can be wrriten as
\be  
\begin{split}
    \langle|F^{\delta}(k)F^{p *}_{\ell}(k)|\rangle=\frac{2\ell+1}{ {A_{\delta}A_{p}}} \int\frac{d \Omega_k}{4\pi} \Big[      \sum_{\ell'} \int \frac{d^3k'}{(2\pi)^3}   P^{\delta p}_{\ell'}(k' ) G^{\delta}({\bf k}-{\bf k}') S^{p*}_{\ell\ell'}({\bf k},{\bf k}')  + N^{\delta p}_{\ell} \Big]
\end{split}
\ee
which is Eq.\ref{eq8s}.

\bibliography{FQinRef}{}

\begin{thebibliography}{}
\expandafter\ifx\csname natexlab\endcsname\relax\def\natexlab#1{#1}\fi
\providecommand{\url}[1]{\href{#1}{#1}}
\providecommand{\dodoi}[1]{doi:~\href{http://doi.org/#1}{\nolinkurl{#1}}}
\providecommand{\doeprint}[1]{\href{http://ascl.net/#1}{\nolinkurl{http://ascl.net/#1}}}
\providecommand{\doarXiv}[1]{\href{https://arxiv.org/abs/#1}{\nolinkurl{https://arxiv.org/abs/#1}}}

\bibitem[{{Abolfathi} {et~al.}(2018){Abolfathi}, {Aguado}, {Aguilar}, {Allende
  Prieto}, {Almeida}, {Ananna}, {Anders}, {Anderson}, {Andrews}, {Anguiano},
  {Arag{\'o}n-Salamanca}, {Argudo-Fern{\'a}ndez}, {Armengaud}, {Ata},
  {Aubourg}, {Avila-Reese}, {Badenes}, {Bailey}, {Balland}, {Barger},
  {Barrera-Ballesteros}, {Bartosz}, {Bastien}, {Bates}, {Baumgarten},
  {Bautista}, {Beaton}, {Beers}, {Belfiore}, {Bender}, {Bernardi}, {Bershady},
  {Beutler}, {Bird}, {Bizyaev}, {Blanc}, {Blanton}, {Blomqvist}, {Bolton},
  {Boquien}, {Borissova}, {Bovy}, {Bradna Diaz}, {Brandt}, {Brinkmann},
  {Brownstein}, {Bundy}, {Burgasser}, {Burtin}, {Busca}, {Ca{\~n}as},
  {Cano-D{\'\i}az}, {Cappellari}, {Carrera}, {Casey}, {Cervantes Sodi}, {Chen},
  {Cherinka}, {Chiappini}, {Choi}, {Chojnowski}, {Chuang}, {Chung}, {Clerc},
  {Cohen}, {Comerford}, {Comparat}, {Correa do Nascimento}, {da Costa},
  {Cousinou}, {Covey}, {Crane}, {Cruz-Gonzalez}, {Cunha}, {da Silva Ilha},
  {Damke}, {Darling}, {Davidson}, {Dawson}, {de Icaza Lizaola}, {de la
  Macorra}, {de la Torre}, {De Lee}, {de Sainte Agathe}, {Deconto Machado},
  {Dell'Agli}, {Delubac}, {Diamond-Stanic}, {Donor}, {Downes}, {Drory}, {du Mas
  des Bourboux}, {Duckworth}, {Dwelly}, {Dyer}, {Ebelke}, {Davis Eigenbrot},
  {Eisenstein}, {Elsworth}, {Emsellem}, {Eracleous}, {Erfanianfar},
  {Escoffier}, {Fan}, {Fern{\'a}ndez Alvar}, {Fernandez-Trincado}, {Fernando
  Cirolini}, {Feuillet}, {Finoguenov}, {Fleming}, {Font-Ribera}, {Freischlad},
  {Frinchaboy}, {Fu}, {G{\'o}mez Maqueo Chew}, {Galbany}, {Garc{\'\i}a
  P{\'e}rez}, {Garcia-Dias}, {Garc{\'\i}a-Hern{\'a}ndez}, {Garma Oehmichen},
  {Gaulme}, {Gelfand}, {Gil-Mar{\'\i}n}, {Gillespie}, {Goddard}, {Gonz{\'a}lez
  Hern{\'a}ndez}, {Gonzalez-Perez}, {Grabowski}, {Green}, {Grier}, {Gueguen},
  {Guo}, {Guy}, {Hagen}, {Hall}, {Harding}, {Hasselquist}, {Hawley}, {Hayes},
  {Hearty}, {Hekker}, {Hernandez}, {Hernandez Toledo}, {Hogg},
  {Holley-Bockelmann}, {Holtzman}, {Hou}, {Hsieh}, {Hunt}, {Hutchinson},
  {Hwang}, {Jimenez Angel}, {Johnson}, {Jones}, {J{\"o}nsson}, {Jullo}, {Khan},
  {Kinemuchi}, {Kirkby}, {Kirkpatrick}, {Kitaura}, {Knapp}, {Kneib},
  {Kollmeier}, {Lacerna}, {Lane}, {Lang}, {Law}, {Le Goff}, {Lee}, {Li}, {Li},
  {Lian}, {Liang}, {Lima}, {Lin}, {Long}, {Lucatello}, {Lundgren}, {Mackereth},
  {MacLeod}, {Mahadevan}, {Maia}, {Majewski}, {Manchado}, {Maraston},
  {Mariappan}, {Marques-Chaves}, {Masseron}, {Masters}, {McDermid}, {McGreer},
  {Melendez}, {Meneses-Goytia}, {Merloni}, {Merrifield}, {Meszaros}, {Meza},
  {Minchev}, {Minniti}, {Mueller}, {Muller-Sanchez}, {Muna}, {Mu{\~n}oz},
  {Myers}, {Nair}, {Nandra}, {Ness}, {Newman}, {Nichol}, {Nidever},
  {Nitschelm}, {Noterdaeme}, {O'Connell}, {Oelkers}, {Oravetz}, {Oravetz},
  {Ort{\'\i}z}, {Osorio}, {Pace}, {Padilla}, {Palanque-Delabrouille},
  {Palicio}, {Pan}, {Pan}, {Parikh}, {P{\^a}ris}, {Park}, {Peirani},
  {Pellejero-Ibanez}, {Penny}, {Percival}, {Perez-Fournon}, {Petitjean},
  {Pieri}, {Pinsonneault}, {Pisani}, {Prada}, {Prakash}, {Queiroz}, {Raddick},
  {Raichoor}, {Barboza Rembold}, {Richstein}, {Riffel}, {Riffel}, {Rix},
  {Robin}, {Rodr{\'\i}guez Torres}, {Rom{\'a}n-Z{\'u}{\~n}iga}, {Ross},
  {Rossi}, {Ruan}, {Ruggeri}, {Ruiz}, {Salvato}, {S{\'a}nchez}, {S{\'a}nchez},
  {Sanchez Almeida}, {S{\'a}nchez-Gallego}, {Santana Rojas}, {Santiago},
  {Schiavon}, {Schimoia}, {Schlafly}, {Schlegel}, {Schneider}, {Schuster},
  {Schwope}, {Seo}, {Serenelli}, {Shen}, {Shen}, {Shetrone}, {Shull}, {Silva
  Aguirre}, {Simon}, {Skrutskie}, {Slosar}, {Smethurst}, {Smith}, {Sobeck},
  {Somers}, {Souter}, {Souto}, {Spindler}, {Stark}, {Stassun}, {Steinmetz},
  {Stello}, {Storchi-Bergmann}, {Streblyanska}, {Stringfellow}, {Su{\'a}rez},
  {Sun}, {Szigeti}, {Taghizadeh-Popp}, {Talbot}, {Tang}, {Tao}, {Tayar},
  {Tembe}, {Teske}, {Thakar}, {Thomas}, {Tissera}, {Tojeiro}, {Tremonti},
  {Troup}, {Urry}, {Valenzuela}, {van den Bosch}, {Vargas-Gonz{\'a}lez},
  {Vargas-Maga{\~n}a}, {Vazquez}, {Villanova}, {Vogt}, {Wake}, {Wang},
  {Weaver}, {Weijmans}, {Weinberg}, {Westfall}, {Whelan}, {Wilcots}, {Wild},
  {Williams}, {Wilson}, {Wood-Vasey}, {Wylezalek}, {Xiao}, {Yan}, {Yang},
  {Ybarra}, {Y{\`e}che}, {Zakamska}, {Zamora}, {Zarrouk}, {Zasowski}, {Zhang},
  {Zhao}, {Zhao}, {Zheng}, {Zheng}, {Zhou}, {Zhu}, {Zinn}, \&
  {Zou}}]{Abolfathi2018}
{Abolfathi}, B., {Aguado}, D.~S., {Aguilar}, G., {et~al.} 2018, \apjs, 235, 42,
  \dodoi{10.3847/1538-4365/aa9e8a}

\bibitem[{{Adams} \& {Blake}(2017)}]{Adams2017}
{Adams}, C., \& {Blake}, C. 2017, \mnras, 471, 839,
  \dodoi{10.1093/mnras/stx1529}

\bibitem[{{Adams} \& {Blake}(2020)}]{Adams2020}
---. 2020, \mnras, 494, 3275, \dodoi{10.1093/mnras/staa845}

\bibitem[{{Alam} {et~al.}(2017){Alam}, {Ata}, {Bailey}, {Beutler}, {Bizyaev},
  {Blazek}, {Bolton}, {Brownstein}, {Burden}, {Chuang}, {Comparat}, {Cuesta},
  {Dawson}, {Eisenstein}, {Escoffier}, {Gil-Mar{\'\i}n}, {Grieb}, {Hand}, {Ho},
  {Kinemuchi}, {Kirkby}, {Kitaura}, {Malanushenko}, {Malanushenko}, {Maraston},
  {McBride}, {Nichol}, {Olmstead}, {Oravetz}, {Padmanabhan},
  {Palanque-Delabrouille}, {Pan}, {Pellejero-Ibanez}, {Percival}, {Petitjean},
  {Prada}, {Price-Whelan}, {Reid}, {Rodr{\'\i}guez-Torres}, {Roe}, {Ross},
  {Ross}, {Rossi}, {Rubi{\~n}o-Mart{\'\i}n}, {Saito}, {Salazar-Albornoz},
  {Samushia}, {S{\'a}nchez}, {Satpathy}, {Schlegel}, {Schneider},
  {Sc{\'o}ccola}, {Seo}, {Sheldon}, {Simmons}, {Slosar}, {Strauss}, {Swanson},
  {Thomas}, {Tinker}, {Tojeiro}, {Maga{\~n}a}, {Vazquez}, {Verde}, {Wake},
  {Wang}, {Weinberg}, {White}, {Wood-Vasey}, {Y{\`e}che}, {Zehavi}, {Zhai}, \&
  {Zhao}}]{Alam2017}
{Alam}, S., {Ata}, M., {Bailey}, S., {et~al.} 2017, \mnras, 470, 2617,
  \dodoi{10.1093/mnras/stx721}

\bibitem[{{Appleby} {et~al.}(2023){Appleby}, {Tonegawa}, {Park}, {Hong}, {Kim},
  \& {Yoon}}]{Appleby2023}
{Appleby}, S., {Tonegawa}, M., {Park}, C., {et~al.} 2023, \apj, 958, 180,
  \dodoi{10.3847/1538-4357/acff68}

\bibitem[{{Beutler} {et~al.}(2019){Beutler}, {Castorina}, \&
  {Zhang}}]{Beutler2019}
{Beutler}, F., {Castorina}, E., \& {Zhang}, P. 2019, \jcap, 2019, 040,
  \dodoi{10.1088/1475-7516/2019/03/040}

\bibitem[{{Bianchi} {et~al.}(2015){Bianchi}, {Gil-Mar{\'\i}n}, {Ruggeri}, \&
  {Percival}}]{Bianchi2015}
{Bianchi}, D., {Gil-Mar{\'\i}n}, H., {Ruggeri}, R., \& {Percival}, W.~J. 2015,
  \mnras, 453, L11, \dodoi{10.1093/mnrasl/slv090}

\bibitem[{{Blake}(2019)}]{Blake2019}
{Blake}, C. 2019, \mnras, 489, 153, \dodoi{10.1093/mnras/stz2145}

\bibitem[{{Blake} {et~al.}(2018){Blake}, {Carter}, \& {Koda}}]{Blake2018}
{Blake}, C., {Carter}, P., \& {Koda}, J. 2018, \mnras, 479, 5168,
  \dodoi{10.1093/mnras/sty1814}

\bibitem[{{Blake} {et~al.}(2010){Blake}, {Brough}, {Colless}, {Couch}, {Croom},
  {Davis}, {Drinkwater}, {Forster}, {Glazebrook}, {Jelliffe}, {Jurek}, {Li},
  {Madore}, {Martin}, {Pimbblet}, {Poole}, {Pracy}, {Sharp}, {Wisnioski},
  {Woods}, \& {Wyder}}]{Blake2010}
{Blake}, C., {Brough}, S., {Colless}, M., {et~al.} 2010, \mnras, 406, 803,
  \dodoi{10.1111/j.1365-2966.2010.16747.x}

\bibitem[{{Blake} {et~al.}(2011){Blake}, {Glazebrook}, {Davis}, {Brough},
  {Colless}, {Contreras}, {Couch}, {Croom}, {Drinkwater}, {Forster}, {Gilbank},
  {Gladders}, {Jelliffe}, {Jurek}, {Li}, {Madore}, {Martin}, {Pimbblet},
  {Poole}, {Pracy}, {Sharp}, {Wisnioski}, {Woods}, {Wyder}, \&
  {Yee}}]{Blake2011}
{Blake}, C., {Glazebrook}, K., {Davis}, T.~M., {et~al.} 2011, \mnras, 418,
  1725, \dodoi{10.1111/j.1365-2966.2011.19606.x}

\bibitem[{{Box} \& {Cox}(1964)}]{Box1964}
{Box}, G. E.~P., \& {Cox}, D.~R. 1964, Journal of the Royal Statistical
  Society., 26, 211–252

\bibitem[{{Campbell} {et~al.}(2014){Campbell}, {Lucey}, {Colless}, {Jones},
  {Springob}, {Magoulas}, {Proctor}, {Mould}, {Read}, {Brough}, {Jarrett},
  {Merson}, {Lah}, {Beutler}, {Cluver}, \& {Parker}}]{Campbell2014}
{Campbell}, L.~A., {Lucey}, J.~R., {Colless}, M., {et~al.} 2014, \mnras, 443,
  1231, \dodoi{10.1093/mnras/stu1198}

\bibitem[{{Carreres} {et~al.}(2023){Carreres}, {Bautista}, {Feinstein},
  {Fouchez}, {Racine}, {Smith}, {Amenouche}, {Aubert}, {Dhawan}, {Ginolin},
  {Goobar}, {Gris}, {Lacroix}, {Nuss}, {Regnault}, {Rigault}, {Robert},
  {Rosnet}, {Sommer}, {Dekany}, {Groom}, {Sravan}, {Masci}, \&
  {Purdum}}]{Carreres2023}
{Carreres}, B., {Bautista}, J.~E., {Feinstein}, F., {et~al.} 2023, \aap, 674,
  A197, \dodoi{10.1051/0004-6361/202346173}

\bibitem[{{Carrick} {et~al.}(2015){Carrick}, {Turnbull}, {Lavaux}, \&
  {Hudson}}]{Carrick2015}
{Carrick}, J., {Turnbull}, S.~J., {Lavaux}, G., \& {Hudson}, M.~J. 2015,
  \mnras, 450, 317, \dodoi{10.1093/mnras/stv547}

\bibitem[{{Davis} \& {Scrimgeour}(2014)}]{Davis2014}
{Davis}, T.~M., \& {Scrimgeour}, M.~I. 2014, \mnras, 442, 1117,
  \dodoi{10.1093/mnras/stu920}

\bibitem[{{Djorgovski} \& {Davis}(1987)}]{Djorgovski1987}
{Djorgovski}, S., \& {Davis}, M. 1987, \apj, 313, 59, \dodoi{10.1086/164948}

\bibitem[{{Dressler} {et~al.}(1987){Dressler}, {Lynden-Bell}, {Burstein},
  {Davies}, {Faber}, {Terlevich}, \& {Wegner}}]{Dressler1987}
{Dressler}, A., {Lynden-Bell}, D., {Burstein}, D., {et~al.} 1987, \apj, 313,
  42, \dodoi{10.1086/164947}

\bibitem[{{Dupuy} {et~al.}(2019){Dupuy}, {Courtois}, \& {Kubik}}]{Dupuy2019}
{Dupuy}, A., {Courtois}, H.~M., \& {Kubik}, B. 2019, \mnras, 486, 440,
  \dodoi{10.1093/mnras/stz901}

\bibitem[{{Erdo{\v{g}}du} {et~al.}(2006){Erdo{\v{g}}du}, {Lahav}, {Huchra},
  {Colless}, {Cutri}, {Falco}, {George}, {Jarrett}, {Jones}, {Macri}, {Mader},
  {Martimbeau}, {Pahre}, {Parker}, {Rassat}, \& {Saunders}}]{Erdogdu2006}
{Erdo{\v{g}}du}, P., {Lahav}, O., {Huchra}, J.~P., {et~al.} 2006, \mnras, 373,
  45, \dodoi{10.1111/j.1365-2966.2006.11049.x}

\bibitem[{{Feldman} {et~al.}(1994){Feldman}, {Kaiser}, \&
  {Peacock}}]{Feldman1994}
{Feldman}, H.~A., {Kaiser}, N., \& {Peacock}, J.~A. 1994, \apj, 426, 23,
  \dodoi{10.1086/174036}

\bibitem[{{Foreman-Mackey} {et~al.}(2013){Foreman-Mackey}, {Hogg}, {Lang}, \&
  {Goodman}}]{Foreman-Mackey2013}
{Foreman-Mackey}, D., {Hogg}, D.~W., {Lang}, D., \& {Goodman}, J. 2013, \pasp,
  125, 306, \dodoi{10.1086/670067}

\bibitem[{{Gorski} {et~al.}(1989){Gorski}, {Davis}, {Strauss}, {White}, \&
  {Yahil}}]{Gorski1989}
{Gorski}, K.~M., {Davis}, M., {Strauss}, M.~A., {White}, S. D.~M., \& {Yahil},
  A. 1989, \apj, 344, 1, \dodoi{10.1086/167771}

\bibitem[{{Hartlap} {et~al.}(2007){Hartlap}, {Simon}, \&
  {Schneider}}]{Hartlap2007}
{Hartlap}, J., {Simon}, P., \& {Schneider}, P. 2007, \aap, 464, 399,
  \dodoi{10.1051/0004-6361:20066170}

\bibitem[{{Hinton}(2016)}]{ChainConsumer}
{Hinton}, S.~R. 2016, The Journal of Open Source Software, 1, 00045,
  \dodoi{10.21105/joss.00045}

\bibitem[{{Hong} {et~al.}(2019){Hong}, {Staveley-Smith}, {Masters}, {Springob},
  {Macri}, {Koribalski}, {Jones}, {Jarrett}, {Crook}, {Howlett}, \&
  {Qin}}]{Hong2019}
{Hong}, T., {Staveley-Smith}, L., {Masters}, K.~L., {et~al.} 2019, \mnras, 487,
  2061, \dodoi{10.1093/mnras/stz1413}

\bibitem[{{Howlett}(2019)}]{Howlett2019}
{Howlett}, C. 2019, \mnras, 487, 5209, \dodoi{10.1093/mnras/stz1403}

\bibitem[{{Howlett} {et~al.}(2015{\natexlab{a}}){Howlett}, {Manera}, \&
  {Percival}}]{Howlett2015a}
{Howlett}, C., {Manera}, M., \& {Percival}, W.~J. 2015{\natexlab{a}}, Astronomy
  and Computing, 12, 109, \dodoi{10.1016/j.ascom.2015.07.003}

\bibitem[{{Howlett} {et~al.}(2015{\natexlab{b}}){Howlett}, {Manera}, \&
  {Percival}}]{Howlett2015bs}
---. 2015{\natexlab{b}}, {L-PICOLA: Fast dark matter simulation code},
  Astrophysics Source Code Library.
\newblock \doeprint{1507.004}

\bibitem[{{Howlett} {et~al.}(2015{\natexlab{c}}){Howlett}, {Ross}, {Samushia},
  {Percival}, \& {Manera}}]{Howlett2015}
{Howlett}, C., {Ross}, A.~J., {Samushia}, L., {Percival}, W.~J., \& {Manera},
  M. 2015{\natexlab{c}}, \mnras, 449, 848, \dodoi{10.1093/mnras/stu2693}

\bibitem[{{Howlett} {et~al.}(2022){Howlett}, {Said}, {Lucey}, {Colless}, {Qin},
  {Lai}, {Tully}, \& {Davis}}]{Howlett2022}
{Howlett}, C., {Said}, K., {Lucey}, J.~R., {et~al.} 2022, \mnras, 515, 953,
  \dodoi{10.1093/mnras/stac1681}

\bibitem[{{Howlett} {et~al.}(2017){Howlett}, {Staveley-Smith}, {Elahi}, {Hong},
  {Jarrett}, {Jones}, {Koribalski}, {Macri}, {Masters}, \&
  {Springob}}]{Howlett2017}
{Howlett}, C., {Staveley-Smith}, L., {Elahi}, P.~J., {et~al.} 2017, \mnras,
  471, 3135, \dodoi{10.1093/mnras/stx1521}

\bibitem[{{Hunter}(2007)}]{Hunter2007}
{Hunter}, J.~D. 2007, Computing in Science Engineering, 9, 90,
  \dodoi{10.1109/MCSE.2007.55}

\bibitem[{{Huterer} {et~al.}(2017){Huterer}, {Shafer}, {Scolnic}, \&
  {Schmidt}}]{Huterer2017}
{Huterer}, D., {Shafer}, D.~L., {Scolnic}, D.~M., \& {Schmidt}, F. 2017, \jcap,
  5, 015, \dodoi{10.1088/1475-7516/2017/05/015}

\bibitem[{{Johnson} {et~al.}(2014){Johnson}, {Blake}, {Koda}, {Ma}, {Colless},
  {Crocce}, {Davis}, {Jones}, {Magoulas}, {Lucey}, {Mould}, {Scrimgeour}, \&
  {Springob}}]{Johnson2014}
{Johnson}, A., {Blake}, C., {Koda}, J., {et~al.} 2014, \mnras, 444, 3926,
  \dodoi{10.1093/mnras/stu1615}

\bibitem[{{Lai} {et~al.}(2023){Lai}, {Howlett}, \& {Davis}}]{Lai2023}
{Lai}, Y., {Howlett}, C., \& {Davis}, T.~M. 2023, \mnras, 518, 1840,
  \dodoi{10.1093/mnras/stac3252}

\bibitem[{Lewis {et~al.}(2000)Lewis, Challinor, \& Lasenby}]{Lewis:1999bs}
Lewis, A., Challinor, A., \& Lasenby, A. 2000, \apj, 538, 473,
  \dodoi{10.1086/309179}

\bibitem[{{Lilow} {et~al.}(2024){Lilow}, {Ganeshaiah Veena}, \&
  {Nusser}}]{Lilow2024}
{Lilow}, R., {Ganeshaiah Veena}, P., \& {Nusser}, A. 2024, arXiv e-prints,
  arXiv:2404.02278, \dodoi{10.48550/arXiv.2404.02278}

\bibitem[{{Linder} \& {Cahn}(2007)}]{Linder2007}
{Linder}, E.~V., \& {Cahn}, R.~N. 2007, Astroparticle Physics, 28, 481,
  \dodoi{10.1016/j.astropartphys.2007.09.003}

\bibitem[{{Ma} {et~al.}(2012){Ma}, {Branchini}, \& {Scott}}]{Ma2012}
{Ma}, Y.-Z., {Branchini}, E., \& {Scott}, D. 2012, \mnras, 425, 2880,
  \dodoi{10.1111/j.1365-2966.2012.21671.x}

\bibitem[{{Macaulay} {et~al.}(2013){Macaulay}, {Wehus}, \&
  {Eriksen}}]{Macaulay2013}
{Macaulay}, E., {Wehus}, I.~K., \& {Eriksen}, H.~K. 2013, \prl, 111, 161301,
  \dodoi{10.1103/PhysRevLett.111.161301}

\bibitem[{{Magoulas} {et~al.}(2012){Magoulas}, {Springob}, {Colless}, {Jones},
  {Campbell}, {Lucey}, {Mould}, {Jarrett}, {Merson}, \&
  {Brough}}]{Magoulas2012}
{Magoulas}, C., {Springob}, C.~M., {Colless}, M., {et~al.} 2012, \mnras, 427,
  245, \dodoi{10.1111/j.1365-2966.2012.21421.x}

\bibitem[{{McDonald} \& {Roy}(2009)}]{McDonald2009}
{McDonald}, P., \& {Roy}, A. 2009, \jcap, 8, 020,
  \dodoi{10.1088/1475-7516/2009/08/020}

\bibitem[{{Nusser} \& {Davis}(1995)}]{Nusser1995}
{Nusser}, A., \& {Davis}, M. 1995, \mnras, 276, 1391,
  \dodoi{10.1093/mnras/276.4.1391}

\bibitem[{{Okumura} {et~al.}(2014){Okumura}, {Seljak}, {Vlah}, \&
  {Desjacques}}]{Okumura2014}
{Okumura}, T., {Seljak}, U., {Vlah}, Z., \& {Desjacques}, V. 2014, \jcap, 2014,
  003, \dodoi{10.1088/1475-7516/2014/05/003}

\bibitem[{{Park}(2000)}]{Park2000}
{Park}, C. 2000, \mnras, 319, 573, \dodoi{10.1046/j.1365-8711.2000.03886.x}

\bibitem[{{Park} \& {Park}(2006)}]{Park2006}
{Park}, C.-G., \& {Park}, C. 2006, \apj, 637, 1, \dodoi{10.1086/498258}

\bibitem[{{Peacock}(1999)}]{Peacock1999}
{Peacock}, J.~A. 1999, {Cosmological Physics}

\bibitem[{{Peebles}(1980)}]{Peebles1980}
{Peebles}, P.~J.~E. 1980, {The large-scale structure of the universe}

\bibitem[{{Peebles}(1993)}]{Peebles1993}
---. 1993, {Principles of Physical Cosmology}, \dodoi{10.1515/9780691206721}

\bibitem[{{Pike} \& {Hudson}(2005)}]{Pike2005}
{Pike}, R.~W., \& {Hudson}, M.~J. 2005, \apj, 635, 11, \dodoi{10.1086/497359}

\bibitem[{{Planck Collaboration} {et~al.}(2020){Planck Collaboration},
  {Aghanim}, {Akrami}, {Ashdown}, {Aumont}, {Baccigalupi}, {Ballardini},
  {Banday}, {Barreiro}, {Bartolo}, {Basak}, {Battye}, {Benabed}, {Bernard},
  {Bersanelli}, {Bielewicz}, {Bock}, {Bond}, {Borrill}, {Bouchet}, {Boulanger},
  {Bucher}, {Burigana}, {Butler}, {Calabrese}, {Cardoso}, {Carron},
  {Challinor}, {Chiang}, {Chluba}, {Colombo}, {Combet}, {Contreras}, {Crill},
  {Cuttaia}, {de Bernardis}, {de Zotti}, {Delabrouille}, {Delouis}, {Di
  Valentino}, {Diego}, {Dor{\'e}}, {Douspis}, {Ducout}, {Dupac}, {Dusini},
  {Efstathiou}, {Elsner}, {En{\ss}lin}, {Eriksen}, {Fantaye}, {Farhang},
  {Fergusson}, {Fernandez-Cobos}, {Finelli}, {Forastieri}, {Frailis},
  {Fraisse}, {Franceschi}, {Frolov}, {Galeotta}, {Galli}, {Ganga},
  {G{\'e}nova-Santos}, {Gerbino}, {Ghosh}, {Gonz{\'a}lez-Nuevo}, {G{\'o}rski},
  {Gratton}, {Gruppuso}, {Gudmundsson}, {Hamann}, {Handley}, {Hansen},
  {Herranz}, {Hildebrandt}, {Hivon}, {Huang}, {Jaffe}, {Jones}, {Karakci},
  {Keih{\"a}nen}, {Keskitalo}, {Kiiveri}, {Kim}, {Kisner}, {Knox},
  {Krachmalnicoff}, {Kunz}, {Kurki-Suonio}, {Lagache}, {Lamarre}, {Lasenby},
  {Lattanzi}, {Lawrence}, {Le Jeune}, {Lemos}, {Lesgourgues}, {Levrier},
  {Lewis}, {Liguori}, {Lilje}, {Lilley}, {Lindholm}, {L{\'o}pez-Caniego},
  {Lubin}, {Ma}, {Mac{\'\i}as-P{\'e}rez}, {Maggio}, {Maino}, {Mandolesi},
  {Mangilli}, {Marcos-Caballero}, {Maris}, {Martin}, {Martinelli},
  {Mart{\'\i}nez-Gonz{\'a}lez}, {Matarrese}, {Mauri}, {McEwen}, {Meinhold},
  {Melchiorri}, {Mennella}, {Migliaccio}, {Millea}, {Mitra},
  {Miville-Desch{\^e}nes}, {Molinari}, {Montier}, {Morgante}, {Moss}, {Natoli},
  {N{\o}rgaard-Nielsen}, {Pagano}, {Paoletti}, {Partridge}, {Patanchon},
  {Peiris}, {Perrotta}, {Pettorino}, {Piacentini}, {Polastri}, {Polenta},
  {Puget}, {Rachen}, {Reinecke}, {Remazeilles}, {Renzi}, {Rocha}, {Rosset},
  {Roudier}, {Rubi{\~n}o-Mart{\'\i}n}, {Ruiz-Granados}, {Salvati}, {Sandri},
  {Savelainen}, {Scott}, {Shellard}, {Sirignano}, {Sirri}, {Spencer},
  {Sunyaev}, {Suur-Uski}, {Tauber}, {Tavagnacco}, {Tenti}, {Toffolatti},
  {Tomasi}, {Trombetti}, {Valenziano}, {Valiviita}, {Van Tent}, {Vibert},
  {Vielva}, {Villa}, {Vittorio}, {Wandelt}, {Wehus}, {White}, {White},
  {Zacchei}, \& {Zonca}}]{Planck2018}
{Planck Collaboration}, {Aghanim}, N., {Akrami}, Y., {et~al.} 2020, \aap, 641,
  A6, \dodoi{10.1051/0004-6361/201833910}

\bibitem[{{Qin}(2021)}]{Qin2021b}
{Qin}, F. 2021, Research in Astronomy and Astrophysics, 21, 242,
  \dodoi{10.1088/1674-4527/21/10/242}

\bibitem[{{Qin} {et~al.}(2019{\natexlab{a}}){Qin}, {Howlett}, \&
  {Staveley-Smith}}]{Qin2019b}
{Qin}, F., {Howlett}, C., \& {Staveley-Smith}, L. 2019{\natexlab{a}}, \mnras,
  487, 5235, \dodoi{10.1093/mnras/stz1576}

\bibitem[{{Qin} {et~al.}(2018){Qin}, {Howlett}, {Staveley-Smith}, \&
  {Hong}}]{Qin2018}
{Qin}, F., {Howlett}, C., {Staveley-Smith}, L., \& {Hong}, T. 2018, \mnras,
  477, 5150, \dodoi{10.1093/mnras/sty928}

\bibitem[{{Qin} {et~al.}(2019{\natexlab{b}}){Qin}, {Howlett}, {Staveley-Smith},
  \& {Hong}}]{Qin2019}
---. 2019{\natexlab{b}}, \mnras, 482, 1920, \dodoi{10.1093/mnras/sty2826}

\bibitem[{{Qin} {et~al.}(2023){Qin}, {Parkinson}, {Hong}, \& {Sabiu}}]{Qin2023}
{Qin}, F., {Parkinson}, D., {Hong}, S.~E., \& {Sabiu}, C.~G. 2023, \jcap, 2023,
  062, \dodoi{10.1088/1475-7516/2023/06/062}

\bibitem[{{Qin} {et~al.}(2021){Qin}, {Parkinson}, {Howlett}, \&
  {Said}}]{Qin2021a}
{Qin}, F., {Parkinson}, D., {Howlett}, C., \& {Said}, K. 2021, \apj, 922, 59,
  \dodoi{10.3847/1538-4357/ac249d}

\bibitem[{{Quelle} \& {Maroto}(2020)}]{Quelle2020}
{Quelle}, A., \& {Maroto}, A.~L. 2020, European Physical Journal C, 80, 369,
  \dodoi{10.1140/epjc/s10052-020-7941-7}

\bibitem[{{Said} {et~al.}(2020){Said}, {Colless}, {Magoulas}, {Lucey}, \&
  {Hudson}}]{Said2020}
{Said}, K., {Colless}, M., {Magoulas}, C., {Lucey}, J.~R., \& {Hudson}, M.~J.
  2020, \mnras, 497, 1275, \dodoi{10.1093/mnras/staa2032}

\bibitem[{{Sakia}(1992)}]{Sakia1992}
{Sakia}, R.~M. 1992, Journal of the Royal Statistical Society, 41, 169,
  \dodoi{10.2307/2348250}

\bibitem[{{Scoccimarro}(2015)}]{Scoccimarro2015}
{Scoccimarro}, R. 2015, \prd, 92, 083532, \dodoi{10.1103/PhysRevD.92.083532}

\bibitem[{{Scrimgeour} {et~al.}(2016){Scrimgeour}, {Davis}, {Blake},
  {Staveley-Smith}, {Magoulas}, {Springob}, {Beutler}, {Colless}, {Johnson},
  {Jones}, {Koda}, {Lucey}, {Ma}, {Mould}, \& {Poole}}]{Scrimgeour2016}
{Scrimgeour}, M.~I., {Davis}, T.~M., {Blake}, C., {et~al.} 2016, \mnras, 455,
  386, \dodoi{10.1093/mnras/stv2146}

\bibitem[{{Sellentin} \& {Heavens}(2016)}]{Sellentin2016}
{Sellentin}, E., \& {Heavens}, A.~F. 2016, \mnras, 456, L132,
  \dodoi{10.1093/mnrasl/slv190}

\bibitem[{{Shi} {et~al.}(2024){Shi}, {Zhang}, {Mao}, \& {Gu}}]{Shi2024}
{Shi}, Y., {Zhang}, P., {Mao}, S., \& {Gu}, Q. 2024, \mnras, 528, 4922,
  \dodoi{10.1093/mnras/stae274}

\bibitem[{{Smith}(2009)}]{Smith2009}
{Smith}, R.~E. 2009, \mnras, 400, 851, \dodoi{10.1111/j.1365-2966.2009.15490.x}

\bibitem[{{Springob} {et~al.}(2014){Springob}, {Magoulas}, {Colless}, {Mould},
  {Erdo{\u{g}}du}, {Jones}, {Lucey}, {Campbell}, \& {Fluke}}]{Springob2014}
{Springob}, C.~M., {Magoulas}, C., {Colless}, M., {et~al.} 2014, \mnras, 445,
  2677, \dodoi{10.1093/mnras/stu1743}

\bibitem[{{Strauss} \& {Willick}(1995)}]{Strauss1995}
{Strauss}, M.~A., \& {Willick}, J.~A. 1995, \physrep, 261, 271,
  \dodoi{10.1016/0370-1573(95)00013-7}

\bibitem[{{Turner} \& {Blake}(2023)}]{Turner2023b}
{Turner}, R.~J., \& {Blake}, C. 2023, \mnras, 526, 337,
  \dodoi{10.1093/mnras/stad2713}

\bibitem[{{Turner} {et~al.}(2021){Turner}, {Blake}, \& {Ruggeri}}]{Turner2021}
{Turner}, R.~J., {Blake}, C., \& {Ruggeri}, R. 2021, \mnras, 502, 2087,
  \dodoi{10.1093/mnras/stab212}

\bibitem[{{Turner} {et~al.}(2023){Turner}, {Blake}, \& {Ruggeri}}]{Turner2023}
---. 2023, \mnras, 518, 2436, \dodoi{10.1093/mnras/stac3256}

\bibitem[{{Virtanen} {et~al.}(2020){Virtanen}, {Gommers}, {Oliphant},
  {Haberland}, {Reddy}, {Cournapeau}, {Burovski}, {Peterson}, {Weckesser},
  {Bright}, {van der Walt}, {Brett}, {Wilson}, {Millman}, {Mayorov}, {Nelson},
  {Jones}, {Kern}, {Larson}, {Carey}, {Polat}, {Feng}, {Moore}, {Vand erPlas},
  {Laxalde}, {Perktold}, {Cimrman}, {Henriksen}, {Quintero}, {Harris},
  {Archibald}, {Ribeiro}, {Pedregosa}, {van Mulbregt}, \& {SciPy 1. 0
  Contributors}}]{Virtanen2020}
{Virtanen}, P., {Gommers}, R., {Oliphant}, T.~E., {et~al.} 2020, Nature
  Methods, 17, 261, \dodoi{10.1038/s41592-019-0686-2}

\bibitem[{{Vlah} {et~al.}(2012){Vlah}, {Seljak}, {McDonald}, {Okumura}, \&
  {Baldauf}}]{Vlah2012}
{Vlah}, Z., {Seljak}, U., {McDonald}, P., {Okumura}, T., \& {Baldauf}, T. 2012,
  \jcap, 11, 009, \dodoi{10.1088/1475-7516/2012/11/009}

\bibitem[{{Vlah} {et~al.}(2013){Vlah}, {Seljak}, {Okumura}, \&
  {Desjacques}}]{Vlah2013}
{Vlah}, Z., {Seljak}, U., {Okumura}, T., \& {Desjacques}, V. 2013, \jcap, 2013,
  053, \dodoi{10.1088/1475-7516/2013/10/053}

\bibitem[{{Wang} {et~al.}(2019){Wang}, {Percival}, {Avila}, {Crittenden}, \&
  {Bianchi}}]{wang2018}
{Wang}, M.~S., {Percival}, W.~J., {Avila}, S., {Crittenden}, R., \& {Bianchi},
  D. 2019, \mnras, 786, \dodoi{10.1093/mnras/stz829}

\bibitem[{{Wang} {et~al.}(2021){Wang}, {Peery}, {Feldman}, \&
  {Watkins}}]{YuyuWang2021}
{Wang}, Y., {Peery}, S., {Feldman}, H.~A., \& {Watkins}, R. 2021, \apj, 918,
  49, \dodoi{10.3847/1538-4357/ac0e37}

\bibitem[{{Wang} {et~al.}(2018){Wang}, {Rooney}, {Feldman}, \&
  {Watkins}}]{YuyuWang2018}
{Wang}, Y., {Rooney}, C., {Feldman}, H.~A., \& {Watkins}, R. 2018, \mnras, 480,
  5332, \dodoi{10.1093/mnras/sty2224}

\bibitem[{{Watkins} \& {Feldman}(2015)}]{Watkins2015}
{Watkins}, R., \& {Feldman}, H.~A. 2015, \mnras, 450, 1868,
  \dodoi{10.1093/mnras/stv651}

\bibitem[{{Whitford} {et~al.}(2023){Whitford}, {Howlett}, \&
  {Davis}}]{Whitford2023}
{Whitford}, A.~M., {Howlett}, C., \& {Davis}, T.~M. 2023, \mnras, 526, 3051,
  \dodoi{10.1093/mnras/stad2764}

\bibitem[{{Yamamoto} {et~al.}(2006){Yamamoto}, {Nakamichi}, {Kamino},
  {Bassett}, \& {Nishioka}}]{Yamamoto2006}
{Yamamoto}, K., {Nakamichi}, M., {Kamino}, A., {Bassett}, B.~A., \& {Nishioka},
  H. 2006, \pasj, 58, 93, \dodoi{10.1093/pasj/58.1.93}

\bibitem[{{Zhang} {et~al.}(2017){Zhang}, {Qin}, \& {Wang}}]{Zhang2017}
{Zhang}, Y., {Qin}, F., \& {Wang}, B. 2017, \prd, 96, 103523,
  \dodoi{10.1103/PhysRevD.96.103523}

\end{thebibliography}
\bibliographystyle{aasjournal}



\end{document}